%% file: main.tex
\documentclass[11pt]{article}

\usepackage{fullpage}
\usepackage[T1]{fontenc}
\usepackage[usenames,dvipsnames]{xcolor}
\usepackage{stmaryrd}

\usepackage{mathpazo}
\usepackage{xfrac}
\usepackage{bm}
\usepackage{todonotes}
\usepackage{setspace}
\usepackage{scrextend}
\usepackage{enumitem}

\RequirePackage[colorlinks=true]{hyperref}
\hypersetup{
  linkcolor=[rgb]{0.3,0.3,0.6},
  citecolor=[rgb]{0.2, 0.6, 0.2},
  urlcolor=[rgb]{0.6, 0.2, 0.2}
}
\usepackage{tikz}
\usetikzlibrary{shapes.geometric}
\usetikzlibrary{decorations.markings}

\usepackage{atmacros}
\input{articleMacros}

\input{notationMacros}
\input{localMacros} 
\input{customurlbst/bibmacros}

\onehalfspace

\title{Monotone Classes Beyond VNP}
\author{
	Prerona Chatterjee \thanks{Blavatnik School of Computer Science, Tel Aviv University. Research supported by the Azrieli International Postdoctoral Fellowship, the Israel Science Foundation (grant number 514/20) and the Len Blavatnik and the Blavatnik Family foundation. Part of this work was done as a postdoctoral researcher at the Institute of Mathematics, Czech Academy of Sciences where I was supported by Grant GX19-27871X of the Czech Science Foundation. Part of this work was also done as a PhD student at TIFR, Mumbai (partially supported by a Google PhD fellowship). Email: \texttt{prerona.ch@gmail.com}.}
	\and
	Kshitij Gajjar \thanks{Indian Institute of Technology Jodhpur, Rajasthan, India. Part of this work was done as a postdoc at NUS, Singapore (NUS ODPRT Grant WBS No. R-252-000-A94-133). Email: \texttt{kshitij@iitj.ac.in}.}
	\and
	Anamay Tengse \thanks{Department of Computer Science, University of Haifa, Israel. Research supported by the Israel Science Foundation (grant No. 716/20). Email: \texttt{anamay.tengse@gmail.com}.}
}

\begin{document}
\maketitle

\pagenumbering{gobble}
\begin{abstract}
	In this work, we study the natural monotone analogues of various equivalent definitions of $\VPSPACE$: a well studied class (Poizat 2008, Koiran \& Perifel 2009, Malod 2011, Mahajan \& Rao 2013) that is believed to be larger than $\VNP$.
	We observe that these monotone analogues are not equivalent unlike their non-monotone counterparts, and propose \emph{monotone $\VPSPACE$} ($\mVPSPACE$) to be defined as the monotone analogue of Poizat's definition.
	With this definition, $\mVPSPACE$ turns out to be exponentially stronger than $\mVNP$ and also satisfies several desirable closure properties that the other analogues may not.
	
    Our initial goal was to understand the monotone complexity of \emph{transparent polynomials}, a concept that was recently introduced by \Hrubes{} \& Yehudayoff (2021).
	In that context, we show that transparent polynomials of large sparsity are hard for the monotone analogues of all the known definitions of $\VPSPACE$, except for the one due to Poizat.
\end{abstract}


\newpage

\pagenumbering{arabic}
\setcounter{page}{1}

\section{Introduction}
The aim of algebraic complexity is to classify polynomials in terms of how hard it is to compute them, and the most standard model for computing polynomials is that of an \emph{algebraic circuit}.
An algebraic circuit is a rooted, directed acyclic graph where the leaves are labelled with variables or field constants and internal nodes are labelled with addition $(+)$ or multiplication $(\times)$.
Every node therefore naturally computes a polynomial and the polynomial computed by the root is said to be the polynomial computed by the circuit.
A formal definition can be found in \autoref{sec:prelims}.

The central question in the area is to show super-polynomial lower bounds against algebraic circuits for \emph{explicit} polynomials, or equivalently, to show that $\VP \neq \VNP$: the algebraic analogue of the famed $\P$ vs. $\NP$ question. 
However, proving strong lower bounds against circuits has turned out to be a difficult problem. 
Much of the research therefore naturally focusses on various restricted algebraic models which compute correspondingly structured polynomials.

One such syntactic restriction is that of \emph{monotonicity}, where the models are not allowed to use any negative constants.
Therefore, trivially, monotone circuits always compute polynomials with only non-negative coefficients.
Such polynomials are called \emph{monotone polynomials}.
We denote the class of all polynomials that are efficiently computable by monotone algebraic circuits by $\mVP$.
Also note that any monomial computed during intermediate computation in a monotone circuit can never get cancelled out, making it a fairly weak model.
As a result, several strong lower bounds are known against monotone circuits.

\paragraph*{Lower bounds in the monotone setting}

There has been a long line of classical works that prove lower bounds against monotone algebraic circuits \cite{Sch76, SS77, SS80, JS82, KZ86, Gas87}.
The most well-known among these, is the result of Jerrum \& Snir \cite{JS82}, where they showed exponential lower bounds against monotone circuits for many polynomial families including the Permanent ($\Perm_n$).
In particular, they showed that every monotone algebraic circuit computing the $n^2$-variate $\Perm_n$ must have size at least $2^{\Omega(n)}$.
A few of the more recent works on monotone lower bounds include \cite{RY11, GS12, CKR20}.

Additionally, many separations that are believed to be true in the general setting have actually been proved to be true in the monotone setting \cite{SS77, HY16, Y19, S20}.
Most remarkably, Yehudayoff \cite{Y19} showed an exponential separation between the computational powers of the monotone analogues of $\VP$ and $\VNP$.
We denote these classes by $\mVP$ (\autoref{defn:mVP}) and $\mVNP$ (\autoref{defn:mVNP}) respectively.

Another line of work in this setting tries to understand the power of non-monotone computational models while computing monotone polynomials.
Valiant \cite{V80}, in his seminal paper, showed that there is a family of monotone polynomials which can be computed by polynomial sized non-monotone algebraic circuits such that any monotone algebraic circuit computing them must have exponential size.
More recent works \cite{HY13, CDM21, CDGM22, CGM22} have shown even stronger separations between the relative powers of monotone and non-monotone models while computing monotone polynomials.

\paragraph*{Newton polytopes, transparency and monotone complexity}
Returning briefly to the general setting, an interesting conjecture relating the algebraic complexity of a bivariate polynomial to its geometric property is the `Tau-conjecture' (also written as $\tau$-conjecture).
The Newton polytope of an $n$-variate polynomial $f$, denoted by $\polytope(f)$, is the convex hull in $\R^n$ of the \emph{exponent vectors} of the monomials in the support of $f$.
Recently, \Hrubes{} \& Yehudayoff \cite{HY21} proposed studying the \emph{Shadows of Newton polytopes} (projections to two-dimensional planes) as an approach to refute the $\tau$-conjecture for Newton polygons made by Koiran, Portier, Tavenas \& \Thomasse{} \cite{KPTT15}. 

Informally, the $\tau$-conjecture for Newton polygons \cite{KPTT15} states that if $f$ is a bivariate polynomial that can be written as an $s$-sum of $r$-products of $p$-sparse polynomials, then its Newton polygon has at most $\poly(s,r,p)$ vertices.
A formal definition of Newton polytopes and the $\tau$-conjecture for Newton polygons can be found in \autoref{sec:prelims}.

This conjecture is fairly strong, and it implies, among other things, that $\VP \neq \VNP$.
However, observe that the Newton polygon retains no information about the coefficients of the polynomial.
Since the algebraic complexity of polynomials is believed to be heavily dependent on coefficients (for example the determinant ($\Det_n$) is efficiently computable by algebraic circuits and this is expected to not be the case for $\Perm_n$, even though they have the same set of monomials), the $\tau$-conjecture for Newton polygons is believed to be false.

The approach suggested by \Hrubes{} \& Yehudayoff~\cite{HY21} used shadows of Newton polytopes as a means to move from the multivariate setting to the bivariate setting, and use polynomials like determinant ($\Det_n$) to refute the conjecture.
The difficulty in this strategy however, is to find a polynomial in $\VP$ that exhibits high \emph{shadow complexity} (maximum number of vertices in its projection), since even when a candidate polynomial is fixed, say $\Det_n$, it is not easy to design a suitable bivariate projection.

As a means to tackle this issue, \Hrubes{} \& Yehudayoff introduced the notion of \emph{transparent polynomials} --- polynomials that can be projected to bivariates in such a way that all of their monomials become vertices of the resulting Newton polygon.
Further, they also gave examples of polynomials with exponentially large sets of monomials that are provably transparent.
Therefore, a proof of any one of these polynomials being in $\VP$ would directly refute the $\tau$-conjecture for Newton polytopes.

Even though \Hrubes{} \& Yehudayoff~\cite{HY21} were not able to actually use this approach to refute the conjecture, they used the notions of shadows \& transparency to come up with yet another method for proving lower bounds against monotone algebraic circuits.
They showed that the monotone circuit complexity of a polynomial is lower bounded by its shadow complexity when the polynomial is transparent.

\begin{theorem}[{\cite[Theorem 2]{HY21}}]
	If $f$ is transparent then every monotone circuit computing $f$ has size at least $\Omega(\abs{\supp(f)})$.
\end{theorem}

As a corollary, they present an $n$-variate polynomial such that any monotone algebraic circuit computing it must have size $\Omega(2^{n/3})$. 

\subsection{Our Contribution}
Here we state our contributions informally; the formal statements can be found in \autoref{sec:contributions-formal}. 
Throughout this work we assume that the underlying field is either the field of real numbers or the field of rational numbers.
The goal of this work is two-fold.

The first goal is to understand how restrictive the notion of transparency is. 
Our search begins with an observation by Yehudayoff~\cite{Y19}, that any lower bound against $\mVP$ depending solely on the support of the hard polynomial, automatically ``lifts'' to $\mVNP$ with the same parameters\footnote{\cite{Y19}: ``If a monotone circuit-size lower bound for $q(\vecx)$ holds also for all polynomials that are equivalent to $q(\vecx)$ then the same lower bound also holds for every $\mVNP$ circuit computing $q(\vecx)$.'' Here $\mVNP$ circuit denotes $\sum_{\vecz \in \set{0,1}^{m}} \ckt(\vecx, z_1, \ldots, z_m)$ where $m = \poly(n)$ and $\ckt(\vecx,\vecz)$ is a monotone algebraic circuit.}.
Since transparency is a property solely of the Newton polytope, and hence of the support of the polynomial, the above observation shows that any transparent polynomial that is non-sparse (has super-polynomially large support) is hard to compute even for $\mVNP$.
However, we suspect that transparency is an even stronger property. 
Therefore, a natural question for us is whether there are even larger classes of monotone polynomials that do not contain non-sparse, transparent polynomials.

This brings us to the second goal of this work --- studying monotone models of computation that can possibly compute polynomials outside even $\mVNP$. 
Classes larger than $\VNP$ had not been defined in the monotone world prior to this work.
We therefore turn to the literature in the non-monotone setting.
Here, $\VPSPACE$ is a well studied class \cite{P08,KP09b,M11,MR13} that is believed to be strictly larger than $\VNP$.
Interestingly there are multiple definitions of $\VPSPACE$, resulting from varied motivations, all of which are known to be essentially equivalent \cite{M11, MR13}.
We study the natural monotone analogues of these definitions and show that unlike the non-monotone setting, the powers of the different resulting models vary greatly.
This allows us to then analyse if the technique of \Hrubes{} \& Yehudayoff also works against monotone classes that are possibly larger than $\mVNP$.

The following figure succinctly describes some of our main results.

\begin{figure}[ht]
	\begin{center} \tikzstyle{vis}=[rectangle, thick, draw=lightgray, fill=orange!10]
	\tikzstyle{vis2}=[rectangle, thick, draw=lightgray, fill=green!10]
	\begin{tikzpicture}
		\node[vis] (ccvp) [] at (5,6) {\large{$\mVP$}};
		\node[vis] (ccsuccabp) [] at (7.2,6) {\large{$\mathsf{msuccABP}$}};
		\node[vis] (ccvnp) [] at (9.5,6) {\large{$\mVNP$}};
		\node[vis] (ccvpquant) [] at (11.7,6) {\large{$\mVP_{\mathsf{quant}}$}};
		\node[vis] (ccvpsumprod) [] at (14.4,6) {\large{$\mVP_{\mathsf{sum, prod}}$}};
		\node[vis2, ellipse] (ccvpproj) [] at (17.2,6) {\large{$\mVP_{\mathsf{proj}}$}};
		
		\begin{scope}[decoration={markings, mark=at position 0.7 with {\arrow[scale=2,>=stealth]{>}}}, darkgray]
		\draw[postaction={decorate}] (ccvp) to[out=30,in=150] (ccvnp);
		\draw[densely dashed, postaction={decorate}] (ccvnp) to[out=0,in=180] (ccvpquant);
		\draw[densely dashed, postaction={decorate}] (ccvpquant) to[out=0,in=180] (ccvpsumprod);
		\draw[densely dashed, postaction={decorate}] (ccvpsumprod) to[out=0,in=180] (ccvpproj);
		\draw[densely dashed, postaction={decorate}] (ccvp) to[out=0,in=180] (ccsuccabp);
		\draw[densely dashed, postaction={decorate}] (ccsuccabp) to[out=-27,in=-153] (ccvnp);
		\draw[postaction={decorate}] (ccvpquant) to[out=27,in=153] (ccvpproj);
		\end{scope}
		
		\node[minimum size=0pt] [] at (7.2,7.2) {\cite{Y19}};
		\node[minimum size=0pt] [] at (14.4,7.3) {\autoref{thm:QuantMonVP-neq-VP-proj}};
		\node[minimum size=0pt] [] at (8.4,5.2) {\autoref{thm:fullMonABP-inside-mVNP}};
	\end{tikzpicture}
	\caption{Nodes represent classes of polynomial families; $A\dashrightarrow B \equiv A\subseteq B$ and $A\longrightarrow B \equiv A\subsetneq B$. Transparent polynomials are hard for all models corresponding to orange, rectangular nodes.}
	\label{fig:classes_inclusion}
	\end{center}
\end{figure}
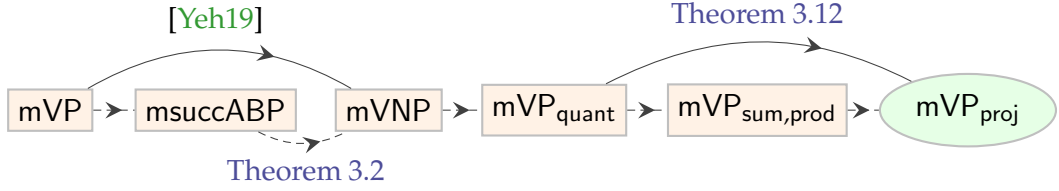

In \autoref{fig:classes_inclusion}, the node labels refer to the following classes of polynomial families that have degree-$\poly(n)$ and $\poly(n)$-complexity under the corresponding models. 
\begin{itemize}
	\item \textsf{msuccABP} - monotone succinct ABPs (\autoref{defn:monotone-succinct-ABPs}),
	\item $\mathsf{mVP}_{\mathsf{quant}}$ - quantified monotone circuits (\autoref{defn:quantified-monotone-algebraic-circuits}),
	\item $\mathsf{mVP}_{\summation,\production}$ - monotone circuits with summation and production gates (\autoref{defn:monotone-circuits-summation-production}),
	\item $\mathsf{mVP}_{\mathsf{proj}}$ - monotone circuits with projection gates (\autoref{defn:monotone-algebraic-circuits-with-projection}).
\end{itemize}
The orange, rectangular nodes denote the classes in which sparsity of transparent polynomials in it is bounded by a constant factor of the size of the smallest $\mathcal{M}$ computing it, if $\mathcal{M}$ is the computational model corresponding to the class (\autoref{thm:transparency-hard-for-QuantMonVP}).

An interesting point to note here is that there is an exponential separation between $\mVP_{\mathsf{quant}}$ and $\mVP_{\mathsf{proj}}$, which means that at least one of the inclusions: $\mVP_{\mathsf{quant}}$ to $\mVP_{\mathsf{sum,prod}}$, and $\mVP_{\mathsf{sum,prod}}$ to $\mVP_{\mathsf{proj}}$ is strict with an exponential separation.

Additionally, we show the following two statements about $\mVP_{\mathsf{quant}}$.
\begin{itemize}
	\item $\mVP_{\mathsf{quant}} = \mVNP$ if and only if homogeneous components of polynomials in $\mVP_{\mathsf{quant}}$ are contained in $\mVP_{\mathsf{quant}}$ (\autoref{cor:mVPquant-equal-mVNP-equiv-condn}).
	In particular, we show that homogeneous polynomials in $\mVP_{\mathsf{quant}}$ are also in $\mVNP$ (\autoref{thm:hom-mVPquant-in-mVNP}). 
	\item $\mVP_{\mathsf{quant}} = \mVP_{\mathsf{sum,prod}}$ if and only if quantified monotone circuits are closed under compositions (\autoref{obs:mVPquant-equal-mVPsumprod-equiv-condn}).
\end{itemize}

Finally, we also show that the homogeneous components of polynomials in $\mVP_{\mathsf{proj}}$ are in $\mVP_{\mathsf{proj}}$ (\autoref{thm:mVPSPACE-ub-for-hom-components}).
This property, along with the fact that $\Perm_n \in \mVP_{\mathsf{proj}}$ (\autoref{thm:Perm-mvpspace-ub}), is the reason we propose ``monotone $\VPSPACE$'' ($\mVPSPACE$) to be defined as the class of polynomial families that are efficiently computable by monotone circuits with projection gates (without any restriction on degree).    

\subsection{Organization of the paper}

We begin in \autoref{sec:prelims} with formal definitions for all the models of computation that we will be using.
Next, we define the monotone analogues of the various definitions of $\VPSPACE$, and outline our results about them in \autoref{sec:contributions-formal}.
The proofs of our results are discussed in \autoref{sec:succinct-monotone-abp}, \autoref{sec:quantified-circuits}, \autoref{sec:monotone-circuits-summation-production} and \autoref{sec:monotone-circuits-projection}.
We conclude with \autoref{sec:conclusion}, where we discuss some important open threads from our work.

\section{Preliminaries}\label{sec:prelims}

We shall use the following notation for the rest of the paper.
\begin{itemize}
	\item We use the standard shorthand $[n] = \set{1,2,\ldots,n}$.
	
	\item We use boldface letters like $\vecx,\vecz,\vece$ to denote tuples/sets of variables or constants, individual members are expressed using indexed version of the usual symbols: $\vece = (e_1,e_2,\ldots,e_n)$, $\vecx=\set{x_1,\ldots,x_n}$.
	We also use $\abs{\vecy}$ to denote the size/length of a vector $\vecy$.

	For vectors $\vecx$ and $\vece$ of the same length $n$, we use the shorthand $\vecx^{\vece} $ to denote the monomial $x_1^{e_1} x_2^{e_2} \cdots x_n^{e_n} $. 

	\item For a polynomial $f(\vecx)$, we denote by $\deg(f)$ the degree of $f$ in $\vecx$.

	\item For a polynomial $f(\vecx)$ and a monomial $m = \vecx^{\vece} $, we refer to the coefficient of $m$ in $f$ by $\coeff_f(m) $.
	The support $\supp(f)$ of a polynomial $f$ is given by $\set{m : \coeff_f(m) \neq 0}$, and the \emph{sparsity} of a polynomial is the size of its support, $\abs{\supp(f)}$.

	\item For any polynomial $f(\vecx)$ and any $k \leq \deg(f)$, we denote by $\hom_k(f)$ the $k$-th homogeneous degree component of $f$ in terms of $\vecx$.
	That is, if $f(\vecx) = f_0(\vecx) + \ldots + f_{\deg(f)}(\vecx)$ where $f_k(\vecx)$ is a homogeneous polynomial of degree $k$ in $\vecx$, then $\hom_k(f) = f_k$.

	\item The permanent of an $n \times n$ symbolic matrix shall be denoted by $\Perm_n$ and is defined as $\Perm_n = \sum_{\sigma \in S_n} \prod_{i=1}^{n} x_{i,\sigma(i)}$, where $S_n$ is the set of all permutations of $[n]$.
	
	\item We use $\set{f_n}$ to denote families of polynomials indexed by $\N$. All complexity classes are defined in terms of asymptotic properties of ``polynomials'' and are technically sets of such polynomial families. Sometimes however, this technicality is ignored for the sake of brevity, especially when the analogous statement about polynomial families is obvious.  
\end{itemize}

\begin{definition}[Algebraic circuits]\label{defn:algebraic-circuits}
	An algebraic circuit is a directed acyclic graph with leaves (nodes with in-degree zero) labelled by formal variables and constants from the field, and other nodes labelled by addition $(+)$ and multiplication $(\times)$ have in-degree 2.
	
	The leaves compute their labels, and every other node computes the operation it is labelled by, on the polynomials along its incoming edges.
	A node of out-degree zero is called the output of the circuit, and the circuit is said to compute the polynomial computed by the output gate.

	In case there are multiple output gates, the circuit is said to be \emph{multi-output}, and computes a set of polynomials.

	The \emph{size} of a circuit, $\ckt$, denoted by $\size(\ckt)$, is the number of nodes in the graph.

	An algebraic circuit over $\Q$ or $\R$ is said to be \emph{monotone}, if all the constants appearing in it are non-negative.
\end{definition}

\begin{definition}[Algebraic Branching Programs (ABPs)]\label{defn:ABP}
	An \emph{algebraic branching program} is specified by a layered graph where each edge is labelled by an affine linear form and the first and the last layer have one vertex each, called the ``source'' and the ``sink'' vertex respectively. 
	The polynomial computed by an ABP is equal to the sum of the weights of all paths from the start vertex to the end vertex in the ABP, where the weight of a path is equal to the product of the labels of all the edges on it. 
	
	The width of a layer in an ABP is the number of vertices in it and the width of an ABP is the width of the layer that has the maximum number of vertices in it.
	The size of an ABP is the number of vertices in it. 
\end{definition}

\begin{definition}[Newton polytopes]\label{defn:newton-polytopes}
For a polynomial $f(\vecx)$, its Newton polytope $\polytope(f) \subseteq \R^n$, is defined as the convex hull of the \emph{exponent vectors} of the monomials in its support.
\[
	\polytope(f) := \operatorname{conv}\inparen{\set{ \vece : \vecx^{\vece} \in \supp(f)}}	
\]

A point $\vece \in \polytope(f)$ is said to be a \emph{vertex}, if it cannot be written as a convex combination of \emph{other} points in $\polytope(f)$.
We denote the set of all vertices of a polytope $\calP$ using $\operatorname{vert}(\calP)$.
\end{definition}

\begin{conjecture}[\texorpdfstring{$\tau$}{Tau} conjecture for Newton polytopes \cite{KPTT15}]
	Suppose $f(x,y)$ is a bivariate polynomial that can be written as $\sum_{i \in [s]} \prod_{j \in [r]} T_{i,j}(x,y)$, where each $T_{i,j}$ has sparsity at most $p$.
	Then the Newton polygon of $f$ has $\poly(s,r,p)$ vertices.
\end{conjecture}

\paragraph*{Basic monotone classes}

\begin{definition}[Monotone $\VP$ ($\mVP$)]\label{defn:mVP}
	A family $\set{f_n}$ of monotone polynomials is said to be in $\mVP$, if there exists a constant $c \in \N$ such that for all large $n$, $f_n$ depends on at most $n^c $ variables, has degree at most $n^c $, and is computable by a monotone algebraic circuit of size at most $n^c $.
\end{definition}
	
\begin{definition}[Monotone $\VNP$ ($\mVNP$)]\label{defn:mVNP}
	A family $\set{f_n}$ of monotone polynomials is said to be in $\mVNP$, if there exists a constant $c \in \N$, and an $m$-variate family $\set{g_m} \in \mVP$ with $m,\size(g_m) \leq n^c$, such that for all large enough $n$, $f_n $ satisfies the following.
	\[
		  f_n(\vecx) = \sum_{\veca \in \set{0,1}^{\abs{\vecy}}} g_{m}(\vecx,\vecy = \veca) \qedhere
	\]
\end{definition}
An expression of the above form is alternatively called \emph{an exponential sum} computing $f_n$.

\subsection*{Various definitions of \textsf{VPSPACE}}

Koiran \& Perifel~\cite{KP09b, KP09} were the first to define $\VPSPACE$ as the class of polynomials (of degree that is potentially exponential in the number of underlying variables) whose coefficients can be computed in $\PSPACE/\poly$, and $\VPSPACE_\text{b}$ to be the polynomials in $\VPSPACE$ that have degree bounded by a polynomial in the number of underlying variables.
They showed that if $\VP \neq \VPSPACE_\text{b}$ then either $\VP \neq \VNP$ or $\P/\poly \neq \PSPACE/\poly$.

Later, Poizat~\cite{P08} gave an alternate definition that does not rely on any boolean machinery, but instead uses a new type of gate called a \emph{projection gate}. 

\begin{definition}[Projection gates \cite{P08}]\label{defn:projection}
	A \emph{projection} gate is a \emph{unary} gate that is labelled by a variable $z$ and a constant $b \in \set{0,1}$, denoted by $\project{z}{b}$.
	It returns the partial evaluation of its input polynomial, at $z=b$, that is, $\project{z}{b}(f(z,\vecx)) = f(b,\vecx)$.
\end{definition}

Poizat defined algebraic circuits with projection gates and then defined $\VPSPACE$ to be the class of polynomial families that are efficiently computable by this model.
Poizat showed\footnote{The work of Poizat is written in French, Malod~\cite{M11} provides an alternate exposition of some of the main results in English.} that this definition is equivalent to that of Koiran \& Perifel.

\begin{definition}[Algebraic circuits with projection gates \cite{P08}]\label{defn:algebraic-circuits-with-projection}
	An \emph{algebraic circuit with projection gates} is an algebraic circuit (\autoref{defn:algebraic-circuits}) in which the internal nodes can also be projection gates (\autoref{defn:projection}), in addition to $+$ or $\times$.

	The size of an algebraic circuit with projection gates is the number of nodes in the underlying graph.
\end{definition}

Adding to Poizat's work, Malod~\cite{M11} characterized $\VPSPACE$ using exponentially large \emph{algebraic branching programs (ABPs)} that are \emph{succinct}.
Malod's work defines the \emph{complexity} of an ABP as the size of the smallest algebraic circuit that encodes its graph --- outputs the corresponding edge label when given the two endpoints as input.
An $n$-variate ABP is then said to be \emph{succinct}, if its complexity is $\poly(n)$.
 
\begin{definition}[Succinct ABPs {\cite{M11}}]\label{defn:succinct-ABPs}
	A \emph{succinct ABP} over the $n$ variables $\vecx = \set{x_1,\ldots,x_n}$ is a triple $(B,\vecs,\vect)$ with $\abs{\vecs} = \abs{\vect} = r$, where
	\begin{itemize}
		\item $\vecs$ is the label of the source vertex, and $\vect$ is the label of the sink(target) vertex.
		\item $B(\vecu,\vecv,\vecx)$ is an algebraic circuit that describes a directed acyclic graph $G_B$ on the vertex set $\set{0,1}^r$ in the following way.
		For any two vertices $\veca,\vecb \in \set{0,1}^r $, the output $B(\vecu = \veca,\vecv = \vecb,\vecx)$ is the label of the edge from $\veca$ to $\vecb$ in the ABP. 
	\end{itemize}
	The polynomial computed by the ABP is the sum of polynomials computed along all $\vecs$ to $\vect$ paths in $G_B$; where each path computes the product of the labels of the constituent edges.
	
	The size of the circuit $B$ is said to be the \emph{complexity} of the succinct ABP.
	The number of vertices $2^r$ is the \emph{size} of the succinct ABP, and the length of the longest $\vecs$ to $\vect$ path is called the \emph{length} of the succinct ABP.
\end{definition}

In the same work~\cite{M11}, Malod alternatively characterized $\VPSPACE$ using an interesting algebraic model that resembles \emph{(totally) quantified boolean formulas} that are known to characterize $\PSPACE$.
This model, which we refer to as ``quantified algebraic circuits'', is defined using special types of projection gates called \emph{summation} and \emph{production} gates.

\begin{definition}[Summation and Production gates \cite{M11}]\label{defn:summation-production}
	\emph{Summation} and \emph{production} gates are unary gates that are labelled by a variable $z$, and are denoted by $\summation_z $ and $\production_z$ respectively.
	A summation gate returns the sum of the $(z=0)$ and $(z=1)$ evaluations of its input, and a production gate returns the product of those evaluations.
	That is, $\summation_z(f(z,\vecx)) = f(0,\vecx) + f(1,\vecx) $, and $\production_z(f(z,\vecx)) = f(0,\vecx) \cdot f(1,\vecx) $.

	We sometimes use $\summation_{\set{z_1,\ldots,z_k}}$ to refer to the nested expression $\summation_{z_1}\cdots\summation_{z_k}$ (similarly for $\production$); it can be checked that the order does not matter here.
\end{definition}

A quantified algebraic circuit has the form $\quant^{1}_{z_1} \quant^{2}_{z_2} \cdots \quant^{m}_{z_m} \ckt(\vecx,\vecz) $, where each $\quant^i $ is a summation or a production, and $\ckt(\vecx,\vecz)$ is a usual algebraic circuit.
\begin{definition}[Quantified Algebraic Circuits \cite{M11}]\label{defn:quantified-algebraic-circuits}
	A quantified algebraic circuit is an algebraic circuit that has the form,
	\[
		\quant^{(1)}_{z_1} \quant^{(2)}_{z_2} \cdots \quant^{(m)}_{z_m} \ckt(\vecx,\vecz),
	\] 
	where $\abs{\vecz} = m$, $\quant^{(i)} \in \set{\summation,\production}$ for each $i \in [m]$, and $\ckt$ is an algebraic circuit.
	The size of such a quantified algebraic circuit is $m + \size(\ckt)$.
\end{definition}

Finally, Mahajan \& Rao~\cite{MR13} defined algebraic analogues of small space computation (e.g. $\textsf{L}$, $\textsf{NL}$) using the notion of \emph{width} of an algebraic circuit.
They use their definitions to import some relationships known in the boolean world to the algebraic world (e.g, they show $\textsf{VL} \subseteq \VP$).
They further show that their definition of uniform polynomially-bounded-space computation coincides with that of \textsf{uniform}-$\VPSPACE$ as defined by Koiran \& Perifel \cite{KP09b}.

We now narrow our focus to the definitions due to Poizat~\cite{P08} and Malod~\cite{M11}.
We choose these definitions because they are algebraic in nature, and have fairly natural monotone analogues.
We elaborate a bit more about this decision in \autoref{sec:boolean-definitions}.

\begin{remark*}
It should be noted that all the above-mentioned definitions of $\VPSPACE$ allow for the polynomial families to have large degree --- as high as $\exp(\poly(n))$.
The main focus of our work, however, is to compare the monotone analogues of these models with $\mVP$ and $\mVNP$.
Since the latter classes only contain low-degree polynomials, we will only work with polynomials of degree $\poly(n)$, or $\VPSPACE_\text{b}$ as defined in \cite{KP09b}, unless mentioned otherwise.
\end{remark*}

\section{Monotone analogues of \textsf{VPSPACE}, and our contributions}\label{sec:contributions-formal}

We now define monotone analogues for the various definitions of $\VPSPACE$ outlined in the previous section, and compare the powers of the resulting monotone models/classes.

\subsection{Monotone succinct ABPs} 
We first consider the natural monotone analogue of the definition due to Malod \cite{M11} which uses succinct algebraic branching programs (\autoref{defn:succinct-ABPs}).

Malod showed that every family $\set{f_n}$ in $\VPSPACE$ can be computed by $2^{\poly(n)}$ sized ABPs that have \emph{complexity} $\poly(n)$.
Recall that the complexity of a succinct ABP is the size of the smallest algebraic circuit that encodes its graph.

We therefore define monotone succinct ABPs as ABPs that can be succinctly described by \emph{monotone} algebraic circuits of size $\poly(n)$.
However, this restriction forces that if the monomial $\vecx^{\vece}$ appears in any edge-label $(\veca,\vecb)$, then it also appears in the label of $(\bar{1},\bar{1})$.
Therefore, self-loops are inevitably present in succinct ABPs in the monotone setting.
To handle this, we additionally allow the \emph{length} of the ABP, say $\ell$, to be predefined\footnote{It is not hard to see that the analogous definition in the non-monotone setting is equivalent to Malod's definition (\autoref{defn:succinct-ABPs}). This is essentially because of the connection to Iterated Matrix Multiplication.} so that now the polynomial computed by the ABP can be defined to be the sum of polynomials computed by all $\vecs$ -- $\vect$ paths of length at most $\ell$. 

\begin{definition}[Monotone Succinct ABPs]\label{defn:monotone-succinct-ABPs}
	A \emph{monotone succinct ABP} over the $n$ variables $\vecx = \set{x_1,\ldots,x_n}$ is a four tuple $(B,\vecs,\vect,\ell)$ with $\abs{\vecs} = \abs{\vect} = r$, where
	\begin{itemize}
		\item $\ell$ is the \emph{length} of the ABP.
		\item $\vecs$ is the label of the source vertex, and $\vect$ is the label of the sink (target) vertex.
		\item $B(\vecu,\vecv,\vecx)$ is a \emph{monotone} algebraic circuit that describes a directed graph $G_B$ on the vertex set $\set{0,1}^r$ in the following way.
		For any two vertices $\veca,\vecb \in \set{0,1}^r $, the output $B(\vecu = \veca,\vecv = \vecb,\vecx)$ is the label of the edge from $\veca$ to $\vecb$ in the ABP. 
	\end{itemize}
	The polynomial computed by the ABP is the sum of polynomials computed along all $\vecs$ to $\vect$ paths in $G_B$ of length at most $\ell$; where each path computes the product of the labels of the constituent edges.
	
	The size of the circuit $B$ is said to be the \emph{complexity} of the monotone succinct ABP.
	The number of vertices $2^r $ is the \emph{size} of the succinct ABP.
\end{definition}
Note that since $B$ is a monotone algebraic circuit, all the edge-labels in the ABP are monotone polynomials over $\vecx$.
It is also not hard to see that any polynomial $f \in \mVP$ is computable by this model.
If $\ckt$ is the monotone circuit computing $f$, then the monotone succinct ABP computing $f$ is $(\ckt', 0, 1, 1)$ where $\ckt'(u,v,\vecx) = v \cdot \ckt(\vecx)$. 

We show that the computational power of monotone succinct ABPs when computing polynomials of \emph{bounded degree} does not even go beyond $\mVNP$.

\begin{restatable}{theorem}{MonSuccinctABPisMonVNP}\label{thm:fullMonABP-inside-mVNP}
	If a polynomial family $\set{f_n}$ of degree $\poly(n)$ is computable by monotone succinct ABPs of complexity $\poly(n)$, then $\set{f_n} \in \mVNP$.
\end{restatable}

In contrast, Malod~\cite{M11} showed that every family in $\VPSPACE$ admits succinct ABPs of polynomial complexity, and we expect $\VPSPACE_b$ to be a much bigger class than $\VNP$.

A proof of \autoref{thm:fullMonABP-inside-mVNP} can be found in \autoref{sec:succinct-monotone-abp}.
It is not clear to us if the converse of \autoref{thm:fullMonABP-inside-mVNP} is true.
Any obvious attack seems to fail due to the restriction that the circuit encoding the ABP needs to be monotone.

\subsection{Quantified monotone circuits}
As mentioned earlier, Malod \cite{M11} had also characterized the class $\VPSPACE$ using the notion of quantified algebraic circuits (\autoref{defn:quantified-algebraic-circuits}).
We now consider its natural monotone analogue, which we call quantified monotone circuits.

\begin{definition}[Quantified Monotone Algebraic Circuits]\label{defn:quantified-monotone-algebraic-circuits}
	A quantified monotone algebraic circuit has the form
	\[
		\quant^{(1)}_{z_1} \quant^{(2)}_{z_2} \cdots \quant^{(m)}_{z_m} \ckt(\vecx,\vecz)
	\] 
	where $\abs{\vecz} = m$, $\quant^{(i)} \in \set{\summation,\production}$ for each $i \in [m]$, and $\ckt$ is a monotone algebraic circuit.
	The size of the quantified monotone algebraic circuit above is $m + \size(\ckt)$.

	We denote by $\mVP_{\mathsf{quant}}$ the class of all $n$-variate polynomial families of degree $\poly(n)$ that are computable by quantified monotone algebraic circuits of size $\poly(n)$.
\end{definition}

Clearly $\mVNP \subseteq \mVP_{\mathsf{quant}}$.
It is therefore interesting to check if the inclusion is tight.
We show that $\mVNP \neq \mVP_{\mathsf{quant}}$ if and only if there is a family $\set{f_n} \in \mVP_{\mathsf{quant}}$ such that the $k$-th homogeneous component of $f_n$ is not in $\mVP_{\mathsf{quant}}$ for some $n$ and $k \leq \deg(f)$. 

In particular we show the following statement. 

\begin{restatable}{theorem}{HomMonVPquantInMonVNP}~\label{thm:hom-mVPquant-in-mVNP}
	Let $f$ be computable by a quantified monotone circuit of size $s$. 
	If $f$ is homogeneous, then it is expressible as an exponential sum of size at most $O(s \cdot \deg(f))$.
\end{restatable}

Since $\mVNP$ is closed under addition, we get the following as a corollary.

\begin{restatable}{corollary}{MonVPquantEqualMonVNPequivCondition}\label{cor:mVPquant-equal-mVNP-equiv-condn}
	The class $\mVP_{\mathsf{quant}}$ is closed under taking homogeneous components, if and only if, $\mVP_{\mathsf{quant}} = \mVNP$.
	That is,
	\[
		\inparen{\forall f \in \mVP_{\mathsf{quant}}, \forall k \leq \deg(f), \hom_k(f) \in \mVP_{\mathsf{quant}}} \iff \mVNP=\mVP_{\mathsf{quant}}
	\]
\end{restatable}

A proof of \autoref{thm:hom-mVPquant-in-mVNP} and \autoref{cor:mVPquant-equal-mVNP-equiv-condn} can be found in \autoref{sec:quantified-circuits}.

Even though we believe $\mVNP \subsetneq \mVP_{\mathsf{quant}}$, we feel this might be tricky to prove.
The following theorem sheds some light on why that may be the case.

\begin{restatable}{theorem}{ExpSumForQuantifiedMVP} 
	\label{thm:exp-sum-for-quantified-mvp}
	Suppose $f(\vecx)$ is an $n$-variate, degree-$d$ polynomial computed by a quantified monotone circuit of size $s$, which uses $\ell$ summation gates.
	Then for a set of variables $\vecw$ of size at most $d \cdot \ell$, there is a monotone circuit $h(\vecx,\vecw)$ of size at most $d \cdot s$, and a monotone polynomial $A(\vecw)$ such that,
	\begin{equation}\label{eq:almost-vnp-expression}
		f(\vecx) = \sum_{\vecb \in \set{0,1}^{\abs{\vecw}}} A(\vecw = \vecb) \cdot h(\vecx,\vecw = \vecb),
	\end{equation}
	where $A(\vecw)$ potentially has circuit size and degree that is exponential in $n$ and $\ell$.
\end{restatable}

Although the obvious size and degree bounds on $A(\vecw)$ above are exponential, it has a somewhat succinct quantified expression that can be inferred from the proof (given in \autoref{sec:quantified-circuits}).

We now discuss how \autoref{thm:exp-sum-for-quantified-mvp} helps us understand a possible difficulty in separating $\mVP_{\mathsf{quant}}$ from $\mVNP$.

\begin{enumerate}
	\item If the polynomial $A(\vecw)$ from \autoref{thm:exp-sum-for-quantified-mvp} were to have degree and size that is polynomial in $n$, then $\mVP_{\mathsf{quant}}$ would collapse to $\mVNP$.
	Further, since $A$ is free of $\vecx$, its exponential degree and size can be leveraged only for designing coefficients of $f$.
	Moreover, the monotone nature of $A$ and $h$ ensures that $A(\mathbf{1})$ is the largest value, and contributes \emph{equally} to all monomials in the support of $f$, since $\supp(f) = \supp(h(\vecx,\vecw=\mathbf{1}))$.
	\item Another consequence that is quite interesting is the following.
	Suppose there is a different monotone polynomial $B(\vecw)$ of small degree and size that agrees with $A(\vecw)$ on all $\set{0,1}$-inputs, then $f(\vecx) = \sum_{\vecb} B(\vecb) h(\vecx,\vecb)$.
	That is, we can replace $A$ by $B$ in our expression and then $f$ clearly has an efficient `$\mVNP$-expression'.
	
	Thus, any separation between $\mVNP$ and quantified monotone $\VP$ will provide a polynomial $A(\vecw)$ which is hard to compute for $\mVNP$, even as a function over the boolean hypercube; a result that perhaps stands on its own. 
\end{enumerate}

\subsection{Monotone circuits with summation and production gates}
Note that it is unclear if quantified monotone circuits are closed under compositions.

We therefore also consider a model that generalizes quantified monotone circuits and is trivially closed under compositions.
Here summation and production gates are allowed to appear anywhere in the circuit.

\begin{definition}[Algebraic circuits with summation and production gates]\label{defn:monotone-circuits-summation-production}
	An \emph{algebraic circuit with summation and production gates} is an algebraic circuit (\autoref{defn:algebraic-circuits}) in which the internal nodes can also be summation or production gates (\autoref{defn:summation-production}), in addition to $+$ or $\times$.	
	A subset of the variables used by the circuit are marked as \emph{auxiliary}.
	These variables do not appear in the output polynomial(s) of the circuit, and the labels for all the summation and production gates are required to be auxiliary variables.

	The \emph{size} of an algebraic circuit with summation and production gates is the number of nodes in the graph.
	
	An algebraic circuit with summation, production gates is said to be \emph{monotone}, if all the constants appearing in it are non-negative.

	We denote by $\mVP_{\mathsf{sum,prod}}$ the class of all $n$-variate polynomial families of degree $\poly(n)$ that are computable by monotone algebraic circuits with summation and production gates of size $\poly(n)$.
\end{definition}

Note that even in the non-monotone setting this model is clearly as powerful as quantified circuits, but can be simulated by circuits with projection gates.
Again, Malod \cite{M11} showed that quantified circuits and circuits with projection gates are equivalent in power. 
So the class of polynomials efficiently computable by this model is also $\VPSPACE$.

In the monotone setting, however, it is not clear if the power of quantified monotone circuits is the same as that of this model.
In particular, we observe the following.
Here, we mean `closure under compositions' in a strong sense: if $C_1$ and $C_2$ are quantified monotone circuits of size $s_1$ and $s_2$ respectively, then the polynomial computed by their composition to have a quantified monotone circuit of size at most $s_1 + s_2$.

\begin{restatable}[Informal]{observation}{MVPquantequalMVPsumprodEquivCondn}\label{obs:mVPquant-equal-mVPsumprod-equiv-condn}
	Quantified monotone circuits are closed under compositions, if and only if, $\mVP_{\mathsf{quant}} = \mVP_{\mathsf{sum,prod}}$.
\end{restatable}

\autoref{thm:quantMVP-composition} gives a formal statement and its proof can be found in \autoref{sec:quantified-circuits}.

We, however, show that even this seemingly stronger model does not help in computing transparent polynomials.

\begin{restatable}{theorem}{TransparencyLBforQuantCircuits}\label{thm:transparency-hard-for-QuantMonVP}
	Any monotone algebraic circuit with summation and production gates that computes a transparent polynomial $f$, has size at least $\abs{\supp(f)}/4$.
\end{restatable}

This shows that transparent polynomials with large support are hard even for this model.
A proof can be found in \autoref{sec:monotone-circuits-summation-production}.

Recall that one way to refute the $\tau$-conjecture for Newton polygons is to show a transparent polynomial in (non-monotone) $\VP$.
\autoref{thm:transparency-hard-for-QuantMonVP} shows that any transparent polynomial from $\VP$ that refutes the conjecture would also witness a separation between $\VP$ and a class potentially much bigger than $\mVNP$\footnote{That is, the class of bounded degree polynomials computable by monotone algebraic circuits with summation and production gates.}.
Even though stark separations between monotone and non-monotone models are not unheard of \cite{HY13, CDM21}, such a result would also be quite interesting and would further highlight the power of subtractions.

\subsection{Monotone circuits with projection gates} 
Finally, adapting the definition of $\VPSPACE$ due to Poizat (\autoref{defn:algebraic-circuits-with-projection}) \cite{P08}, we define monotone circuits with projection gates.

\begin{definition}[Monotone algebraic circuits with projection gates]\label{defn:monotone-algebraic-circuits-with-projection}
	A \emph{monotone algebraic circuit with projection gates} is an algebraic circuit with projections (\autoref{defn:algebraic-circuits-with-projection}) in which only non-negative constants from the field are allowed to appear as labels of leaves.

	As in \autoref{defn:monotone-circuits-summation-production}, only the \emph{auxiliary} variables can be used as labels for the projection gates.	
	The \emph{size} of a monotone algebraic circuit with projection gates is the number of nodes in the underlying graph.

	We denote by $\mVP_{\mathsf{proj}}$ the class of all $n$-variate polynomials of degree $\poly(n)$ that are computable by size-$\poly(n)$ monotone algebraic circuits with projection gates.
\end{definition}

This model is clearly at least as powerful as monotone circuits with summation and production gates, since $\summation_z = \project{z}{0} + \project{z}{1}$ and $\production_z = \project{z}{0} \times \project{z}{1}$.
It would therefore be interesting to show a separation between the power of the two models.

Even though we are unable to do that, we show that monotone circuits with projection gates are indeed more powerful than quantified monotone circuits, with a $2^{\Omega(\sqrt{m})}$ separation. 

\begin{restatable}{theorem}{QuantCircuitsVsProjectionCircuits}\label{thm:QuantMonVP-neq-VP-proj}
The polynomial family $\set{\Perm_n}$ can be computed by monotone circuits with projection gates of size $O(n^3)$, but quantified monotone circuits computing it must have size $2^{\Omega(n)} $.
\end{restatable}

Finally we show that $\mVP_{\mathsf{proj}}$ is closed under taking homogeneous components.

\begin{restatable}{theorem}{mVPSPACEubForHomComponents}\label{thm:mVPSPACE-ub-for-hom-components}
	Suppose $f$ is computed by a size $s$ monotone circuit with projections.
	Then for any $k \leq \deg(f)$, $\hom_k(f)$ has a monotone circuit with projections of size $O(k^2 \cdot s)$.
\end{restatable}

Proof sketches of \autoref{thm:QuantMonVP-neq-VP-proj} and \autoref{thm:mVPSPACE-ub-for-hom-components} can be found in \autoref{sec:monotone-circuits-projection}.

\subsection{Defining Monotone \textsf{VPSPACE} ($\mVPSPACE$)}



We propose the following definition for $\mVPSPACE$.

\begin{definition}[Monotone $\VPSPACE$]\label{defn:monotone-vpspace}
	A family of polynomials $\set{f_n}$ is said to be in $\mVPSPACE$ if for all large $n$, $f_n$ is computable by a monotone algebraic circuit with projection gates (\autoref{defn:monotone-algebraic-circuits-with-projection}) of size $\poly(n)$.
	
	Further if $\set{f_n}$ has degree $\poly(n)$, then it is said to be in $\mVPSPACE_b$.
\end{definition}

That is, we define $\mVPSPACE_b := \mVP_{\mathsf{proj}}$ and define $\mVPSPACE$ along the same lines, but without the restriction on the degree being bounded (since $\VPSPACE$ does not impose any restrictions on degree).
Some of our reasons for this choice are as follows.

Firstly, being a complexity class, $\mVPSPACE_b$ should be closed under (monotone) affine projections, i.e. setting a few variables to monotone affine polynomials.
All of $\mVP_{\mathsf{quant}}$, $\mVP_{\mathsf{sum,prod}}$ and $\mVP_{\mathsf{proj}}$ have this property.

Further, as $\mVP$ and $\mVNP$ are closed under taking homogeneous components, it is desirable for a more powerful class to also have this property.
Even if $\mVP_{\mathsf{quant}}$ satisfies this, it would not lead to a larger class (\autoref{cor:mVPquant-equal-mVNP-equiv-condn}).
Also, it is not clear $\mVP_{\mathsf{sum,prod}}$ is closed under homogenization, while $\mVP_{\mathsf{proj}}$ is (\autoref{thm:mVPSPACE-ub-for-hom-components}).

Finally, we believe that having $\Perm_n \in \mVP_{\mathsf{proj}}$ is an interesting property that further strengthens the case for $\mVP_{\mathsf{proj}}$ being the definition for $\mVPSPACE_b$.

\section{Monotone succinct algebraic branching programs}\label{sec:succinct-monotone-abp}

In this section we discuss the proof of \autoref{thm:fullMonABP-inside-mVNP}. 

\MonSuccinctABPisMonVNP*

\begin{proof}
	Let $\mathcal{A} = (B,\vecs,\vect, \ell)$ be the monotone succinct ABP computing $f$, with $\abs{\vecs} = \abs{\vect} = r$. 
	Then we observe the following.

	\begin{claim}
		If $\ell > 1$, then $\ell \leq \deg(f)+2$.
	\end{claim}
	\begin{proof}
		Let $b(\vecu,\vecv,\vecx)$ be the \emph{monotone} $(2r+n)$-variate polynomial computed by the circuit $B$.
		Due to the monotonicity of $B$, for any $\vece \in \N^n$ we have that if the monomial $\vecx^{\vece}$ appears in any edge-label $(\veca,\vecb)$, then it also appears in the label of $(\bar{1},\bar{1})$.
		Therefore, $\deg_{\vecx}(B(\veca,\vecb,\vecx)) \leq \deg_{\vecx}(B(\bar{1},\bar{1},\vecx)) $ for all $\veca,\vecb$.
		Similarly, $\deg_{\vecx}(B(\vecs,\vecb,\vecx)) \leq \deg_{\vecx}(B(\vecs,\bar{1},\vecx))$ and $\deg_{\vecx}(B(\veca,\vect,\vecx)) \leq \deg_{\vecx}(B(\bar{1},\vect,\vecx))$ for all $\veca,\vecb$.
		This shows that if $\ell > 1$, then 
		\[
			\deg(f) = \deg(B(\vecs,\bar{1},\vecx) \cdot B(\bar{1},\bar{1},\vecx)^{\ell - 2} \cdot B(\bar{1},\vect,\vecx)) \geq \ell - 2. \qedhere
		\]
	\end{proof}

	As a result of the above claim, for $d = \deg(f)$, we have the following.
	\begin{align*}
		f(\vecx) 
		&= B(\vecs,\vect,\vecx) + \sum_{j = 1}^{d + 1} \inparen{\text{sum of $\vecs$--$\vect$ paths through $j$ intermediate vertices}}\\
		&= B(\vecs,\vect,\vecx) + \sum_{j = 1}^{d + 1} \inparen{\sum_{\veca_1,\ldots,\veca_j \in \set{0,1}^r} B(\vecs,\veca_1,\vecx) \cdot \inparen{\prod_{k = 1}^{j-1} B(\veca_k,\veca_{k+1},\vecx) } \cdot B(\veca_{j},\vect,\vecx)}\\
		&= B(\vecs,\vect,\vecx) +\\
		& \quad \sum_{\veca_1,\ldots,\veca_{d+1} \in \set{0,1}^r} \sum_{j = 1}^{d+1} 2^{-r(d+1-j)} \inparen{B(\vecs,\veca_1,\vecx) \cdot \inparen{\prod_{k = 1}^{j-1} B(\veca_k,\veca_{k+1},\vecx) } \cdot B(\veca_{j},\vect,\vecx)}.
	\end{align*}

	\noindent This can be rewritten as follows.
	\[
		\sum_{\veca_1,\ldots,\veca_{d+1}} \inparen{2^{-r(d+1)}B(\vecs,\vect,\vecx) + \sum_{j = 1}^{d+1} 2^{-r(d+1-j)} B(\vecs,\veca_1,\vecx) \inparen{\prod_{k = 1}^{j-1} B(\veca_k,\veca_{k+1},\vecx) } B(\veca_{j},\vect,\vecx)}
	\]
	This is clearly a poly-sized exponential sum as $d = \poly(n)$ and $B$ is a monotone circuit of size $\poly(n)$.
\end{proof}

\section{Quantified monotone circuits}\label{sec:quantified-circuits}

\subsection{Computing homogeneous polynomials}

\HomMonVPquantInMonVNP*

\begin{proof}

Let $d = \deg(f)$, and let $\ckt$ be a quantified monotone circuit computing $f$, that uses exactly $k$ production gates.
We can then assume that, 
\[
	\ckt(\vecx) = \summation_{\vecy_0} \production_{z_1} \summation_{\vecy_1} \production_{z_2} \cdots \summation_{\vecy_{k-1}} \production_{z_k} \summation_{\vecy_{k}} g(\vecx,\vecy,\vecz),
\] 
without loss of generality, by using some empty $\vecy_j$s whenever necessary.
Note that the $\vecy_j$s are sets of variables, whereas each of the $z_j$s are single variables.

We now prove the statement in two steps.
First, we use the homogeneity of $f$, and the monotonicity of the quantified circuit, to show that $k \leq \log(d)$.

\begin{claim}
	$k \leq \log d$
\end{claim}

\begin{proof}
	For each $i \in [k]$, let $g_i(z_i,\vecx,\vecw_i) = \summation_{\vecy_i} \production_{z_{i+1}} \summation_{\vecy_{i+1}} \cdots \summation_{\vecy_{k}} g(\vecx,\vecy,\vecz)$.
	Here $\vecw_i$ denotes all the auxiliary variables that are alive after `$i$ rounds' of quantifiers.
	Further, let $h_i(\vecx,\vecw_i) = \production_{z_i} g_i(z_i,\vecx,\vecw_i)$.

	Now, $f(\vecx) = \summation_{\vecy_0} h_1(\vecx,\vecy_0)$, and it is homogeneous.
	Therefore, since $h_1$ is monotone, it is also homogeneous in $\vecx$ with degree exactly $d$.
	But $\deg_{\vecx}(h_1) = \deg_{\vecx}(\production_{z_1} g_1) = \deg_{\vecx}(g_1(z_1 = 0)) + \deg_{\vecx}(g_1(z_1 = 1))$.
	If we write $g_1(z_1,\vecx,\vecw_1) = g_{1,0}(\vecx,\vecw_1) + z \cdot g_{1,1}(z_1,\vecx,\vecw_1)$, then we have that $g_1(z_1 = 0) = g_{1,0}(\vecx,\vecw_1)$ and $g_1(z_1 = 1) = g_{1,0}(\vecx,\vecw_1) + g_{1,1}(z_1=1,\vecx,\vecw_1)$.
	Since $h_1$ is homogeneous in $\vecx$ and $g_1$ is monotone in all the variables, this must mean that $\deg_{\vecx}(g_1(z_1 = 0)) = \deg_{\vecx}(g_1(z_1 = 1)) = \deg_{\vecx}(g_1) = d/2$.
	Also, $g_1$ is homogeneous in $\vecx$, and thus we can repeat the same argument for $h_2$, $g_2$, and so on.

	As a result, we see that $\deg(f) = 2^k \cdot \deg_{\vecx}(g)$, and hence $k \leq \log d$.
\end{proof}

We can now make $2^k \leq d$ many copies of the `inner circuit' $g(\vecx,\vecy,\vecz)$, one for each fixing of the $\vecz$ variables.
We then obtain the final exponential sum computing $f$ by using the following `product rule' for summations repeatedly.
\[
	(\summation_{\vecy_1} h_1(\vecx,\vecy_1)) \cdot (\summation_{\vecy_2} h_2(\vecx,\vecy_2)) = \summation_{\widetilde{\vecy}_1,\widetilde{\vecy}_2} (h_1(\vecx,\widetilde{\vecy}_1) \cdot h_2(\vecx,\widetilde{\vecy}_2))
\]

Note that in the above case the two summations are over disjoint sets of variables.
This can easily be ensured in our case, by treating the $\vecy$ variables in each of the $2^k \leq d$ copies as mutually disjoint. 
It is easy to see that the exponential sum has size $O(\size(C),d)$.
\end{proof}

\begin{remark}
	The first step in the above proof extends more or less as it is, to an arbitrary circuit with summation and production gates.
	Thus, any circuit with arbitrary summations and productions \emph{that computes a homogeneous polynomial} can be assumed to not contain any production gates, with a polynomial blow-up in size.
	
	However, this does not directly give an efficient exponential sum, because of the second step in the above argument.
	It crucially uses the fact that for any summation gate $g$, the number of production gates on a path from $g$ to the root was $O(\log d)$.
	This ensures that no summation gate (or its auxiliary variable) has to be replicated more than $\poly(d)$ times, which is not necessarily true if we start with an arbitrary circuit with summation gates.
\end{remark}

\subsection{Large exponential sums for arbitrary polynomials}

We shall need the following simple observation, which follows from the `product-rule' for summations stated earlier.
\begin{observation}[Product of exponential sums] \label{obs:product-of-exp-sums}
	\[
		\production_z \summation_{\vecy} g(\vecx,\vecy,z) = \summation_{\vecy_0,\vecy_1} \inparen{g(\vecx,\vecy_0,0) \cdot g(\vecx,\vecy_1,1)} \qedhere
	\]
\end{observation}

Let us see a toy case of trivially moving from a quantified expression to an exponential sum, using \autoref{obs:product-of-exp-sums}.
\begin{align*}
	f(x) &= \summation_{y_1} \production_{z_1} \summation_{y_2} \production_{z_2,z_3} \summation_{y_3} g(x,y_1,y_2,y_3,z_1,z_2,z_3)\\
		 &= \summation_{y_1} \production_{z_1} \summation_{y_2} \production_{z_2} \summation_{y_{3,0},y_{3,1}} \inparen{\prod_{a_3 \in \set{0,1}} g(x,y_1,y_2,y_{3,a_3},z_1,z_2,a_3)}\\
	     &= \summation_{y_1} \production_{z_1}  \summation_{y_2, y_{3,(00)},y_{3,(01)},y_{3,(10)},y_{3,(11)}} \inparen{\prod_{a_2,a_3 \in \set{0,1}} g(\ldots,y_{3,(a_2 a_3)},z_1,a_2,a_3)}\\
		&= \summation_{y_1} \summation_{y_{2,\ast},y_{3,\ast\ast\ast}}\inparen{\prod_{a_1,a_2,a_3 \in \set{0,1}} g(x,y_1,y_{2,a_1},y_{3,(a_1 a_2 a_3)},a_1,a_2,a_3)}
\end{align*}
In the last line, each $\ast$ runs over $\set{0,1}$, so there are $1 + 2+ 8 = 11$ auxiliary variables in total.
Note that $y_3$ has $8$ copies, which is due to the $3$ production gates `above' the summation gate labelled by it.
Similarly, $y_2$ has just $2$ copies, while $y_1$ has just one.
Also, if instead of single auxiliary variables $y_2$ and $y_3$ we had sets of auxiliary variables $\vecy_2 $ and $\vecy_3$, nothing much would change.
That is, we would have had $8$ copies of the set $\vecy_3$ and $2$ copies of $\vecy_2$, irrespective of their sizes.

What this shows in general, is that we can trivially move from a quantified expression to an expression which has the form
\[
	f(\vecx) = \summation_{\mathbf{Y}} \prod_{\veca \in \set{0,1}^r} g_\veca(\vecx, \vecy_\veca)
\]
where $\mathbf{Y} = \cup_{\veca} \set{\vecy_\veca}$, $r$ is the number of production gates in the quantified expression, $\abs{\mathbf{Y}}$ is potentially exponential (since the number of copies of some auxiliary variable might be exponential) but $g_\veca(\vecx, \vecy_\veca) = g(\vecx, \vecy = \vecy_\veca, \vecz = \veca)$ for a poly-sized circuit $g(\vecx, \vecy, \vecz)$.

The key observation that allows us to prove \autoref{thm:exp-sum-for-quantified-mvp} is that if $f$ has degree $d$, then the number of copies of each auxiliary variable needed in the outer summation gate is at most $d$.
This is because, due to monotonicity, $\deg_{\vecx}(g_\veca(\vecx, \vecy_\veca)) \neq 0$ for only $d$ many $\veca \in \set{0,1}^r$.

For a formal proof, we introduce a new shorthand.
For a vector $\veca = \set{a_1,a_2,\ldots,a_{\ell}}$ and a number $k \leq \ell$, we use $\veca[:k]$ to denote the \emph{prefix} vector $\set{a_1,a_2,\ldots,a_k}$.
With this new notation, we can express the last line of our toy example in \autoref{sec:quantified-circuits} is as follows.
\[
	f(x) = \summation_{y_1} \summation_{y_{2,\ast},y_{3,\ast\ast\ast}}\inparen{\prod_{\veca \in \set{0,1}^3} g(x,y_1,y_{2,\veca[:1]},y_{3,\veca[:3]},a_1,a_2,a_3)}
\]
We are now ready to prove \autoref{thm:exp-sum-for-quantified-mvp}, which we recall once more.

\ExpSumForQuantifiedMVP*

\begin{proof}
	The first step is to obtain a trivial exponential sum for the quantified expression, as in the discussion above.
	\begin{claim}\label{claim:trivial-exp-sum}
		Suppose $f(\vecx)$ can be expressed as the following quantified circuit.
		\[
			f(\vecx) = \summation_{\vecy_1} \production_{\vecz_1} \summation_{\vecy_2} \production_{\vecz_2} \cdots \production_{\vecz_{k}} \summation_{\vecy_{k+1}} g(\vecx,\vecy_1,\ldots,\vecy_{k+1},\vecz_1,\ldots,\vecz_{k})
		\]
		Let $m_i = \abs{\vecz_i}$, and further let $M_i = m_1 + m_2 + \cdots + m_i$, for each $i \in [k]$.
		Also, let $\vecy = \vecy_1 \cup \vecy_2 \cup \cdots \cup \vecy_{k+1}$, and $\vecz = \vecz_1 \cup \vecz_2 \cup \cdots \cup \vecz_{k}$
		
		Then $f(\vecx)$ can also be expressed as the following exponential sum.
		\[
			f(\vecx) = \summation_{\mathbf{Y}} \inparen{\prod_{\veca \in \set{0,1}^{M_{k}}} g(\vecx,\vecy_{1},\vecy_{2,\veca[:M_1]},\vecy_{3,\veca[:M_2]},\ldots,\vecy_{k+1,\veca[:M_k]},\vecz = \veca) }
		\]
		Here $\mathbf{Y}$ is a set of all $y$-variables, of size $\inparen{1 + \sum_i 2^{M_i}}$ that is defined as follows.
		\[
			\mathbf{Y} = \bigcup_{\veca \in \set{0,1}^{M_k}} ( \vecy_1 \cup \vecy_{2,\veca[:M1]} \cup \cdots \cup \vecy_{k+1,\veca[:M_k]}) \qedhere
		\]
	\end{claim}
	Even though the claim is fairly verbose, it is easy to verify given the discussion before the lemma, so we will not explicitly prove it.

	As the next step, we shall use the fact that the `inner circuit' $g$ is monotone, to bound the degree of $f$ from below.
	\begin{align*}
		\deg(f) &= \deg_{\vecx}\inparen{ \summation_{\mathbf{Y}} \inparen{\prod_{\veca \in \set{0,1}^{M_{k}}} g(\vecx,\vecy_{1},\vecy_{2,\veca[:M_1]},\ldots,\vecy_{k+1,\veca[:M_k]},\vecz = \veca) } }\\
		\text{($g$ is monotone)} &= \deg_{\vecx}\inparen{\prod_{\veca \in \set{0,1}^{M_{k}}} g(\vecx,\mathbf{1},\vecz = \veca) }\\
				&\geq \sum_{\veca \in \set{0,1}^{M_k}} \deg(g(\vecx,\mathbf{1},\vecz = \veca))
	\end{align*}
	Therefore, since $f$ has degree $d = \deg(f)$, it must be the case that for all but $d$ fixings $\veca$ of $\vecz$, $g(\vecx,\vecy,\veca)$ is a constant in terms of $\vecx$ for any $\set{0,1}$-assignment to the variables in $\vecy$.

	Let $\mathcal{A} := \set{\veca \in \set{0,1}^{M_k} : \deg_{\vecx}\inparen{g(\vecx,\vecb,\veca)} > 0 \text{ for some } \vecb \in \set{0,1}^{\abs{\vecy}} }$, and let $\mathcal{A}_0 := \set{0,1}^{M_k} \setminus \mathcal{A}$.
	We therefore have that $\abs{\mathcal{A}} \leq d$.
	Further, let $\mathbf{Y}_1 := \bigcup_{\veca \in \mathcal{A}} ( \vecy_1 \cup \vecy_{2,\veca[:M1]} \cup \cdots \cup \vecy_{k+1,\veca[:M_k]})$, and let $\mathbf{Y}_0 := \mathbf{Y} \setminus \mathbf{Y}_1 $.
	Note that now $\abs{\mathbf{Y}_1} \leq \abs{\mathcal{A}} \cdot \abs{\vecy} \leq d \cdot m$.
	
	We can now simplify the exponential sum in \autoref{claim:trivial-exp-sum} and finish the proof as follows, where $\vecy_{\veca}$ refers to $(\vecy_1,\vecy_{2,\veca[:M_1]},\cdots,\vecy_{k+1,\veca[:M_k]})$.

	\[
		 = \summation_{\mathbf{Y}} \inparen{\prod_{\veca \in \set{0,1}^{M_{k}}} g(\vecx,\vecy_{\veca},\vecz = \veca) }
		= \summation_{\mathbf{Y}} \inparen{ \inparen{\prod_{\veca \in \mathcal{A}_0} g(\vecx,\vecy_{\veca},\vecz = \veca)} \cdot \inparen{\prod_{\veca \in \mathcal{A}} g(\vecx,\vecy_{\veca},\vecz = \veca)} }\\
	\]
	for appropriate $\vecy_{\veca}$.
	Now this is equal to
	\[
		\summation_{\mathbf{Y}} \inparen{ \inparen{\prod_{\veca \in \mathcal{A}_0} g(\mathbf{0},\vecy_{\veca},\vecz = \veca)} \cdot \inparen{\prod_{\veca \in \mathcal{A}} g(\vecx,\vecy_{\veca},\vecz = \veca)} }
	\]
	since the first term is ``$\vecx$-free''.

	Therefore,
	\begin{align*}
		f(\vecx) &= \summation_{\mathbf{Y}_1,\mathbf{Y}_0} \inparen{ \inparen{\prod_{\veca \in \mathcal{A}_0} g(\mathbf{0},\vecy_{\veca},\vecz = \veca)} \cdot \inparen{\prod_{\veca \in \mathcal{A}} g(\vecx,\vecy_{\veca},\vecz = \veca)} }\\
		\text{(regroup terms)} &= \summation_{\mathbf{Y}_1} \inparen{ \summation_{\mathbf{Y}_0} \inparen{\prod_{\veca \in \mathcal{A}_0} g(\mathbf{0},\vecy_{\veca},\vecz = \veca)}} \cdot \inparen{\prod_{\veca \in \mathcal{A}} g(\vecx,\vecy_{\veca},\vecz = \veca)}\\
		\text{(simplify)}&= \summation_{\mathbf{Y}_1} A(\mathbf{Y}_1) \cdot h(\vecx,\mathbf{Y}_1)
	\end{align*}
	As claimed, the size of $h$ is at most $\abs{\mathcal{A}} \cdot \size(g) \leq d \cdot s$, while $A(\mathbf{Y}_1)$ is a fairly structured polynomial despite its exponential size and degree.
\end{proof}

\begin{remark*}
	Since we are allowed exponential size for $A(\vecw)$ one can always take the multilinear polynomial that agrees with $A$ on the hypercube.
	However, as mentioned towards the end of the proof, we get a monotone polynomial $A(\vecw)$ that is fairly structured.
	This in particular means that an arbitrary multilinear $A$ that is outside $\mVNP$ does not witness the desired separation.
\end{remark*}

\section{Monotone circuits with summation and production gates}\label{sec:monotone-circuits-summation-production}

\subsection{Shadow Complexity of monotone circuits with summation and production gates}
In this section, we begin with a proof of \autoref{thm:transparency-hard-for-QuantMonVP}.
Let us start by recalling the theorem.

\TransparencyLBforQuantCircuits*

This result is an extension of the ideas in the work of \Hrubes{} \& Yehudayoff~\cite{HY21}.
Their argument shows that any bivariate monotone circuit of size $s$ that computes a polynomial with \emph{convexly independent support} outputs a polynomial with support at most $4s$.
They achieve this by keeping track of the largest polygon (in terms of the number of vertices) that one can build using the polynomials computed at all the gates in the circuit.
They then inductively show that no gate (leaf, addition, multiplication) can increase the number of vertices by $4$.
We are able to show the same bound for production and summation gates, by working with a monotone bivariate circuit over $y_1,y_2 $ that is allowed some auxiliary variables $z$ for summations and productions.

An important component of the proof in \cite{HY21} is that if the sum or product of two monotone polynomials is convexly independent, then so are each of the two inputs.
However, allowing for summations and productions means that some monomials that are computed internally could get ``zeroed out''.
In fact, summation and production gates do not quite ``preserve convex dependencies''.
For example, the convexly dependent support $\set{y_1 y_2, y_1 y_2 z, y_1 y_2 z^2} $ when passed through $\summation_z $ produces just $\set{y_1 y_2}$, which is convexly independent.

In order to prove \autoref{thm:transparency-hard-for-QuantMonVP}, one can get around this by working directly with the support projected down to the ``true'' variables, which we call $\vecy$-support in our arguments.
It turns out that summations and productions indeed preserve convex dependencies that are in the $\vecy$ support of the input polynomial.

Before we begin a formal proof, let us recal the concepts of \emph{shadow complexity} and \emph{transparent polynomials}.
\begin{definition}[Shadow complexity {\cite{HY21}}]\label{defn:shadow-complexity}
	For a polynomial $f(x_1,\ldots,x_n)$, its shadow complexity $\sigma(f)$ is defined as follows, where the $\max$ is taken over \emph{linear} maps.
	\begin{equation*}
		\sigma(f) := \max_{L : \R^n \rightarrow \R^2} \abs{\operatorname{vert}(L(\polytope(f)))} \qedhere
	\end{equation*}
\end{definition}
For any $n$, a set of points in $\R^n $ is said to be \emph{convexly independent} if no point in the set can be written as a convex combination of other points from the set.
Note that if a polynomial has \emph{convexly independent support}, then all the monomials in its support correspond to vertices of its Newton polytope.
The following definition is an even stronger condition.
\begin{definition}[Transparent polynomials {\cite{HY21}}]\label{defn:transparent-polynomials}
	A polynomial $f$ is said to be transparent if $\sigma(f) = \abs{\supp(f)}$. 
\end{definition}
The following lemma states that the linear map that witnesses the shadow complexity of a polynomial over the reals, can be assumed to be ``integral'' without loss of generality.
\begin{lemma}[Consequence of {\cite[Lemma 4.2]{HY21}}]\label{lem:real-polynomial-integral-projections}
	Let $f(\vecx) \in \R[\vecx]$ be an $n$-variate polynomial.
	Then there is an $M \in \Z^{2 \times n} $, such that for $L(\vece) := M \cdot \vece$, $\abs{\operatorname{vert}(L(\polytope(f)))} = \sigma(f)$.
\end{lemma}
We also require the following concepts from the work of \Hrubes{} \& Yehudayoff \cite{HY21}.
\begin{definition}[Laurent polynomials and high powered circuits]\label{defn:hp-circuits-laurent}
	A \emph{Laurent polynomial} over the variables $\set{x_1,\ldots,x_n}$ and a field $\F$, is a finite $\F$-linear combination of terms of the form $x_1^{p_1} x_2^{p_2} \cdots x_n^{p_n} $, where $p_1,p_2,\ldots,p_n \in \Z$.
	A high-powered circuit over the variables $\set{x_1,\ldots,x_n}$ and a field $\F$, is an algebraic circuit whose leaves can compute terms like $\alpha x_1^{p_1} x_2^{p_2} \cdots x_n^{p_n} $ for any $\alpha \in \F$ and $\vecp \in \Z^n $.
	In other words, a high-powered circuit can compute an arbitrary Laurent monomial with size $1$; the size of the high-powered circuit is the total number of nodes as usual.
\end{definition}
Using the above definition, we can easily infer the following by replacing each leaf with the corresponding Laurent monomial.
\begin{observation}\label{obs:shadows-hp-circuits}
	Let $f(\vecx)$ be computable by a monotone circuit of size $s$, and suppose $\sigma(f) = k$.
	Then there exists a bivariate Laurent polynomial $P(y_1,y_2)$ that is computable by a high-powered circuit of size $s$, whose Newton polygon has $k$ vertices.
\end{observation}

We will also need the following lemma from \cite{HY21}.
\begin{lemma}[{\cite[Lemma 5.8]{HY21}}]\label{lem:mink-sums-2d}
	Let $A, B \subset \R^2 $ be finite sets, such that $A + B$ is convexly independent.
	Then if $\abs{A} \geq \abs{B}$, then either $\abs{A},\abs{B} \leq 2$ or $\abs{B} = 1$.
\end{lemma}

We now have all the concepts required to prove the main theorem of this section, \autoref{thm:transparency-hard-for-QuantMonVP}.
The following results and their proofs closely follow those in \cite{HY21}.
We reproduce the overlapping parts for the sake of completeness and ease of exposition.

\begin{theorem}[Extension of {\cite[Theorem 5.9]{HY21}}]\label{thm:transparent-summation-production}
	Let $f(y_1,y_2)$ be a monotone Laurent polynomial with convexly independent support, and let $C(y_1,y_2,\vecz)$ be a monotone high-powered circuit with summation and production gates\footnote{All auxiliary variables only appear with non-negative powers in the circuit.}, that computes $f$.
	Then $\size(C) \geq \abs{\supp(f)}/4$.
\end{theorem}
\begin{proof}
	For a multi-set\footnote{We assume that copies of the same set $A \in \calA$ can be referred distinctly.} $\calA$ that contains sets of points in $\R^2$, we define a measure $\mu$ that relates to the ``largest'' convexly independent set that can be constructed using it.
	For a sub-collection $\calB \subseteq \calA$ and a map $v : \calB \rightarrow \R^2 $, the resulting set $\calB(v)$ is defined as follows.
	\begin{equation*}
		\calB(v) := \bigcup_{A \in \calB} (\set{v(A)} + A)
	\end{equation*}
	The measure $\mu$ is then defined as follows.
	\begin{equation}\label{eqn:definition-mu}
		\mu(\calA) := \max_{\calB,v} \set{ \abs{\calB(v)} : \calB(v)\text{ is convexly independent}}
	\end{equation}
	For a Laurent polynomial $g(y_1,y_2,\vecz)$, let $\suppy(g) := \set{(a,b) : \exists \vece, y_1^a y_2^b \vecz^{\vece} \in \supp(g)}$ be its $\vecy$-support.
	Corresponding to the circuit $C(y_1,y_2,\vecz)$ of size $s$, we will consider the collection $\calA$ of $s$ sets, which will be the $\vecy$-supports of the polynomials computed by the $s$ gates.
	The following claim will help us prove the theorem by induction.
	\begin{claim}
	For $\calA' = \calA \cup \set{B}$, and $A_1,A_2 \in \calA $,
	\begin{align}
		\mu(\calA') &\leq \mu(\calA) + \abs{B}, &\label{eq-case:1}\\
		\mu(\calA') &\leq \mu(\calA) + 2 			&\text{if $B = u + A_1 $},\label{eq-case:2}\\
		\mu(\calA') &\leq \mu(\calA) + 4 			&\text{if $B = A_1 \cup A_2 $},\label{eq-case:3}\\
		\mu(\calA') &\leq \mu(\calA) + 4 			&\text{if $B = A_1 + A_2 $},\label{eq-case:4}\\
		\mu(\calA') &\leq \mu(\calA) + 4 			&\text{if $B = A_1 + A' $ for $A' \subseteq A_1$ }\label{eq-case:5}.
	\end{align}
	\end{claim}
	\begin{proof}
	It is trivial to see that \eqref{eq-case:1} holds.
	For \eqref{eq-case:2}, suppose $\calB$ is the subset that achieves $\mu(\calA') > \mu(\calA)$.
	Then $A_1,B \in \calB $ as otherwise one can mimic the contribution of $B$ using $A_1 $; further $v(A_1) \neq v(B) + u$ because otherwise the translations of $A_1 $ and $B$ overlap.
	Now note that $(\set{v(A_1)}+A_1) \cup (\set{v(B)}+B) $ is a convexly independent set of points, and also that $(\set{v(A_1)}+A_1) \cup (\set{v(B)}+B) = \set{v(A_1),v(B)+u} + A_1$.
	Therefore, by \autoref{lem:mink-sums-2d}, we see that $\abs{B} = \abs{A_1} \leq 2$, which finises the proof using \eqref{eq-case:1}.
	For \eqref{eq-case:3}, observe that $\mu(\calA) \leq \mu(\calA \cup {A_1,A_2})$.
	The desired bound then follows by two applications of \eqref{eq-case:2}.
	In \eqref{eq-case:4}, if $B$ is convexly \emph{dependent}, then it cannot contribute to $\mu(\calA')$, so suppose it is.
	Assuming $\abs{A_1} \geq \abs{A_2}$ without loss of generality, by \autoref{lem:mink-sums-2d}, either $\abs{B} \leq \abs{A_1} \cdot \abs{A_2} \leq 4$, or $B = u + A_1 $ for some $u$, and \eqref{eq-case:2} finishes the proof.
	Clearly \eqref{eq-case:4} implies \eqref{eq-case:5}, as its proof does not depend on whether $A_2 \in \calA$, or $A_2 \not\subseteq A_1$.
\end{proof}
We now argue that the polynomial computed at every gate in $C(y_1,y_2,\vecz)$ has convexly independent $\vecy$-support.
Since the $\vecy$-supports of addition and multiplication gates are unions and Minkowski sums of their children respectively, if any of their input is convexly dependent, then so is the output.
For a summation gate $g = \summation_z g'$, $\suppy(g) = \suppy(g')$ using \autoref{lem:support-summation-production}.
For a production gate $g = \production_z g' $, $\suppy(g) = S' + \suppy(g')$ for some $S' \subseteq \suppy(g')$, so any convex dependency in $\suppy(g')$ would transfer to $\suppy(g)$.
Since the output of $C(x,y,\vecz)$ is convexly independent, the above observations imply that each gate $g \in C$ has convexly independent $\suppy(g)$.
Let us now prove the theorem by inductively building the collection $\calA$ with respect to the circuit $C$: a gate is added only after adding all of its children.
When the gate being added is a leaf, then $\mu$ increases by at most $1$ due to \eqref{eq-case:1}.
For an addition gate computing $g$, $\suppy(g) $ is the union of the $(x,y)$-supports of its children; so we can apply \eqref{eq-case:3}.
For a multiplication gate computing $g$, $\suppy(g) $ is the Minkowski sum of the $(x,y)$-supports of its children; so we can use \eqref{eq-case:4}.
For a summation gate that computes $g$, note that its $(x,y)$-support is exactly the same as that of its child (\autoref{obs:support-monotone-projection}); therefore \eqref{eq-case:2} applies.
Finally, for a production gate, we can use \eqref{eq-case:5}, as $\suppy(\production_z g) =  \suppy(g\vert_{z=0}) + \suppy(g\vert_{z=1})$, and $\suppy(g\vert_{z=0}) \subseteq \suppy(g\vert_{z=1}) = \suppy(g)$.
Since the measure $\mu$ increases by at most $4$ in each of the $s$ steps, we have that $\abs{\supp(f)} \leq \mu(\calA) \leq 4s$, as required.
\end{proof}
\noindent The above result then lets us prove \autoref{thm:transparency-hard-for-QuantMonVP}, which we first restate.
\TransparencyLBforQuantCircuits*
\begin{proof}
Let $C$ be a monotone circuit with production and summation gates of size $s$ that computes $f_n$.
Since $f_n(\vecx) \in \R[\vecx] $ is transparent, there exists a matrix $M \in \Z^{2 \times n} $, such that the linear map $L(\vece) = M\vece $, satisfies $\abs{\operatorname{vert}({L(\polytope(f))})} = \abs{\supp(f)}$.
Further, using \autoref{obs:shadows-hp-circuits}, there exists a size-$s$ high-powered monotone circuit with summation and production gates, that computes a Laurent polynomial $P(y_1,y_2)$ which has $\abs{\supp(f)}$ vertices in its Newton polytope.
The bound then easily follows from \autoref{thm:transparent-summation-production}.
\end{proof}

\subsection{Quantified monotone circuits and compositions}

\MVPquantequalMVPsumprodEquivCondn*

Even though this statement appears to be straightforward, formally stating it requires a bit more care.
Doing that yields the following theorem.

\begin{theorem}\label{thm:quantMVP-composition}
	Suppose that for any quantified monotone circuit $\ckt$ of size $s$ with $r$ leaves, and any \emph{multi-output} quantified monotone circuit $\ckt'$ of size $s'$ with $r$ outputs, we have that the polynomial computed by $\ckt \circ \ckt'$ has a quantified monotone circuit of size at most $(s+s')$.

	Then, any \emph{multi-output}, monotone circuit with summation and production gates of size $\tilde{s}$ can be simulated by a \emph{multi-output} quantified monotone circuit of size at most $\tilde{s}$, and hence $\mVP_{\mathsf{quant}} = \mVP_{\mathsf{sum,prod}}$.
	
	The converse is also true.
\end{theorem}

\begin{proof}
	One direction of the implication is clearly true because circuits with (arbitrary) summation and production gates have the stated property by definition.
	
	For the converse, let us assume that quantified monotone circuits have the property.
	We show that this implies that the two models in question have the same power.

	Consider a circuit $\ckt$ of size $s$ with summation and production gates.
	We group the gates in $\ckt$ in ``bands'' numbered from the bottom to the top, in the following way.
	\begin{itemize}\itemsep0pt
		\item The $0$-th band consists only of leaves 
		\item Odd bands consist only of addition or multiplication gates.
		\item Even bands (other than $0$) only consist of summation or production gates.
		\item The gates in band $i$ can have edges incoming from only bands $j \leq i$.
	\end{itemize}
	
	Now, given a circuit $\tilde{\ckt}$ of size $\tilde{s}$ with summation and production gates, we express it as a quantified monotone circuit of size $O(s)$ by inducting on the number of bands in it.

	For the base case, when $\tilde{\ckt}$ has up to two bands, it is already a quantified monotone circuit.

	In general, if $\tilde{\ckt}$ has $2b'$ bands, we look at the circuit formed by bands $2b'$ and $(2b' - 1)$ as a quantified monotone circuit; let its size be $s$.
	By induction, the \emph{multi-output} circuit formed by the bands $0$ to $2b'-2$ can be expressed as a multi-output, quantified monotone circuit of size at most $s' = \tilde{s} - s$, call it $\ckt'$.
	Now from the hypothesis, the composition $\ckt \circ \ckt'$ is also computable by a quantified monotone circuit of size at most $s + s' \leq \tilde{s}$.
\end{proof}

\section{Monotone circuits with projection gates}\label{sec:monotone-circuits-projection}

\subsection{Exponential separation from quantified circuits}

\QuantCircuitsVsProjectionCircuits*

We begin by proving that $\Perm_n \in \mVP_{\mathsf{proj}}$.

\begin{theorem}\label{thm:Perm-mvpspace-ub}
	There is a monotone circuit with projection gates of size $O(n^3)$ that computes $\Perm_n$.
\end{theorem}

\begin{proof}
	We first define a polynomial $P_0$ such that all its monomials contain exactly one $\vecx$-variable from each row.
	\[
		\text{Let } P_0(\vecx,\vecy) := \inparen{\sum_{j=1}^{n} y_{1,j} x_{1,j}} \inparen{\sum_{j=1}^{n} y_{2,j} x_{2,j}} \cdots \inparen{\sum_{j=1}^{n} y_{n,j} x_{n,j}}.
	\]
	Note that $P_0$ has $n^2 $-many auxiliary variables $\vecy$, one attached to each `true' variable $x_{i,j}$.
	We now want to use these to progressively prune the monomials that pick up multiple variables from the $j$th column by projecting the $n$ variables $y_{1,j},\ldots,y_{n,j}$.

	Let $e_1,\ldots,e_n \in \set{0,1}^n $ such that $e_i(k) = 1 \Leftrightarrow i=k$, and define for each $j \in [n]$,
	\begin{equation}
		P_j := \sum_{i \in [n]} \project{y_{1,j}}{e_i(1)}\inparen{\project{y_{2,j}}{e_i(2)}\inparen{\cdots\inparen{\project{y_{n,j}}{e_i(n)}\inparen{P_{j-1}}}}}.
	\end{equation}
	The following claim is now easy to verify.
	\begin{claim}
		For all $j \in [n]$, $P_{j} $ contains all the monomials from $P_{j-1}$ that are supported on exactly one $\vecx$-variable from the $j$th column.
	\end{claim}
	As a result, the monomials in $P_{n} $ are exactly those of the monomials in $\Perm_n $.
	Additionally, for each $j$, the auxiliary variables in $P_j $ are only from the columns $j+1,\ldots,n$; thus $P_n = \Perm_n $.

	The size of our circuit is $O(n^3)$, since $\size(P_0) = O(n^2)$ and $\size(P_j) = \size(P_{j-1}) + O(n^2)$.
	This proves \autoref{thm:Perm-mvpspace-ub}.
\end{proof}

\begin{remark}
	Our upper bound above also implies that any polynomial (family) that can be expressed as the permanent of a monotone matrix of size $\poly(n)$ (called monotone $p$-projection of $\Perm_n$) can also be computed by efficient monotone circuits with projection gates.
	Although $\Perm_n$ is complete for non-monotone $\VNP$, it is \emph{not} the case that all monotone polynomials in $\VNP$ are monotone $p$-projections of $\Perm_n$, as shown by Grochow~\cite{G17}.
\end{remark}

The proof of \autoref{thm:QuantMonVP-neq-VP-proj} now follows from the following simple extension of an observation due to Yehudayoff~\cite{Y19}\footnote{Observation in \cite{Y19}: Let $g(\vecx,z)$ be a monotone polynomial and let $c > 0$. Then for any monomial $m = \vecx^\vece z^j $ in the support of $g$, $\vecx^\vece \in \supp(g,z=c)$.}\label{obs:support-monotone-projection} and the classical lower bound of Jerrum \& Snir \cite{JS82} against monotone algebraic circuits for $\Perm_n$.

\begin{restatable}{lemma}{SupportSummationProduction}\label{lem:support-summation-production}
	Let $f(\vecx)$ be a monotone polynomial whose support cannot be written as a non-trivial product of two sets, and for some monotone polynomial $g(\vecx,\vecz)$, suppose we have $f(\vecx) = \quant^{(1)}_{z_1} \quant^{(2)}_{z_2} \cdots \quant^{(m)}_{z_m} g(\vecx,\vecz) $ with $\quant^{(i)} \in \set{\summation,\production}$ for each $i \in [m]$.
	
	Then $\supp(f(\vecx)) = \supp(g(\vecx,\bar{1}))$.
\end{restatable}

\begin{proof}
	Observe that it is enough to show the statement of the lemma for $m = 1$.
	Therefore, suppose $f(\vecx) = \summation_z g(\vecx,z)$, then $f(\vecx) = g(\vecx,0) + g(\vecx,1)$, and hence $\supp(f) = \supp(g(\vecx,1))$, since $g$ is monotone.
	
	Next, $f(\vecx) = \prod_z g(\vecx,z)$ means that $f(\vecx) = g(\vecx,0) \cdot g(\vecx,1)$.
	As $\supp(f)$ cannot be written as a non-trivial product\footnote{For sets of monomials $A$ and $B$, their product is defined as $A \times B = \set{m \cdot m' : m \in A, m' \in B}$; a non-trivial product is when neither $A$ nor $B$ is just $\set{1}$.} of two sets, and since $g$ is monotone, this must mean that $g(\vecx,0)$ is a constant and $\supp(f(\vecx)) = \supp(g(\vecx,1))$ as claimed.
\end{proof}

Finally, let us complete the proof of \autoref{thm:QuantMonVP-neq-VP-proj}.

\QuantCircuitsVsProjectionCircuits*

\begin{proof}
	Let us assume that there is a quantified monotone circuit of size $s$ computing $\Perm_n$.
	Then,
	\[
		\Perm_n(\vecx) = \quant^{(1)}_{z_1} \quant^{(2)}_{z_2} \cdots \quant^{(m)}_{z_m} f(\vecx,\vecz)	
	\] 
	for some $m \leq s$ and $\quant^{(i)} \in \set{\summation,\production}$ for each $i \in [m]$.

	Note that, by definition, $f(\vecx,\vecz)$ is computable by a monotone algebraic circuit of size at most $s$ and therefore $f(\vecx,\bar{1})$ is computable by a monotone algebraic circuit of size at most $s$.
	On the other hand, by \autoref{lem:support-summation-production}, the support of $f(\vecx,\bar{1})$ is the same as that of $\Perm_n$ since $\Perm_n$ is irreducible.
	The required lower bound now follows from the fact that the $2^{\Omega(n)}$ lower bound proved by Jerrum \& Snir \cite{JS82} for $\Perm_n$ against monotone algebraic circuits, works for any polynomial that has support equal to the support of $\Perm_n$.
\end{proof}

\subsection{Closure under homogenization}

\mVPSPACEubForHomComponents*

\begin{proof}
We show this using the classical argument of `gate replication'.
Given a circuit $\ckt$, we construct another circuit $\ckt'$ that has $(k+1)$ copies of each gate in $\ckt$.
For a gate $g \in \ckt$, the corresponding gates $g_0, g_1, \ldots, g_k$ shall compute $\hom_i([g])$ for each $i \leq k$, where $[g]$ is the polynomial computed at $g$.
Here and throughout the proof, the degree of a polynomial always refers to its degree in the $\vecx$-variables.

The following can now be easily checked, using the fact that $[g]$ is always a monotone polynomial.
\begin{itemize}\itemsep0pt
	\item If $[g]$ is a leaf labelled with a `true' variable $x_i$, then $[g_1] = x_i$ and $[g_i] = 0$ for all other $i$.
	\item If $[g]$ is any other leaf, then $[g_0] = [g]$ and $[g_i] = 0$ for all other $i$.
	\item If $[g] = [u] + [v]$, then $[g_i] = [u_i] + [v_i]$ for all $i$.
	\item If $[g] = \project{z}{b} [u]$, then $[g_i] = \hom_i([g]) = \project{z}{b} \hom_i([u]) = \project{z}{b} [u_i]$.
	\item If $[g] = [u] \times [v]$, then $[g_i] = \sum_{j \leq i} [u_j] \times [v_{i-j}]$, for each $i$.
\end{itemize}

The last case incurs the largest blow-up in size, which adds $O(k^2)$ many gates in $\ckt'$ for one gate in $\ckt$.
This finishes the proof.
\end{proof}

\section{Conclusion}\label{sec:conclusion}

Our work is an attempt at understanding the hardness of transparent polynomials for monotone algebraic models.
We observe that the lower bound of \Hrubes{} \& Yehudayoff~\cite{HY21} extends beyond monotone $\VNP$, and therefore turn to exploring the class $\VPSPACE$ from the non-monotone world.
This exploration reveals that the natural monotone analogues of the multiple equivalent definitions of $\VPSPACE$ have contrasting powers.
Additionally, transparent polynomials turn out to be as hard for some of these analogues as they are for usual monotone circuits.
The following are some interesting open threads from our work.

\begin{itemize}
\item We suspect that transparency is a highly restrictive property, especially for monotone computation.
Therefore, we conjecture that if $f$ is a transparent polynomial being computed by a size-$s$ monotone circuit with projection gates, then $\abs{\supp(f)} \leq 2^{\operatorname{polylog}(s)}.$
It would be interesting (at least to us) to see a proof or a refutation of this conjecture.

An immediate hurdle in extending the techniques in \cite{HY21} (\autoref{thm:transparency-hard-for-QuantMonVP}) to $\mVPSPACE$, is that unlike summations and productions, $0$-projections do not preserve convex dependencies, even if we restrict to the ``true'' variables.

\item Along similar lines, a possibly simpler goal is to show a non-monotone circuit upper bound for a transparent polynomial.
	Since transparency only restricts the support of the polynomial, one is free to choose any real coefficients to ensure that it is in $\VP$ (\autoref{lem:real-polynomial-integral-projections} works for all real polynomials).
	In particular, this brings powerful non-monotone tricks like interpolation into play.
	Among other things, such a result would refute the notoriously open $\tau$-conjecture for Newton polygons.

\item Another question we would like to highlight is separating $\mVNP$ and quantified monotone circuits.
As mentioned in the discussion following \autoref{thm:exp-sum-for-quantified-mvp}, such a separation would yield a (high degree) polynomial that is hard for $\mVNP$ even as a function over the boolean hypercube.
Such a polynomial might be of interest, perhaps, even in the non-monotone setting.
\end{itemize}

\bibliographystyle{customurlbst/alphaurlpp}
\bibliography{references}

\appendix

\section{Definitions of \textsf{VPSPACE} relying on boolean computation}\label{sec:boolean-definitions}
In this section we briefly address why we did not study monotone analogues of the definitions due to Koiran \& Perifel~\cite{KP09b, KP09}, and Mahajan \& Rao~\cite{MR13}.

Koiran \& Perifel define uniform $\VPSPACE$ as the class of families $\set{f_n}$ of $\poly(n)$-variate polynomials of degree at most $2^{\poly(n)} $, such that there is a $\PSPACE$ machine that computes the \emph{coefficient function} of $\set{f_n}$.
Here, the coefficient function of $\set{f_n}$ can be seen to map a pair $(1^n,\vece)$ to the coefficient of $\vecx^{\vece} $ in $f_n $.

Non-uniform $\VPSPACE$ is then defined by replacing $\PSPACE$ by its non-uniform analogue, $\PSPACE/\poly$.
Since there are no monotone analogues of Turing machines, perhaps the only possible monotone analogue of this definition is to insist on the coefficient function being monotone, which results in an absurdly weak class (the ``largest'' monomial will always be present).

Mahajan \& Rao~\cite{MR13} look at the notion of \emph{width} of a circuit --- all gates are assigned heights, such that the height of any gate is \emph{exactly} one larger than the height of its highest child.
The width of the circuit is the maximum number of nodes that have the same height.
They then define $\mathsf{VSPACE}(S(n))$, as the class of families that are computable by circuits of width $S(n)$ and size at most $\max\set{2^{S(n)},\poly(n)} $.

The class uniform $\mathsf{VSPACE}(S(n))$ further requires that the circuits be $\mathsf{DSPACE(S(n))}$-uniform.
Although their non-uniform definition is purely algebraic, it is a bit unnatural for space $S(n) \gg \log{n}$ (as also pointed out in their paper), since such circuits may not even have a $\poly(n)$-sized description.
We therefore do not analyse a monotone analogue for their definition.

\end{document}

%% file: articleMacros.tex
\usepackage{amsthm}
\usepackage{thmtools,thm-restate}

\numberwithin{equation}{section}
\declaretheoremstyle[bodyfont=\it,qed=\qedsymbol]{noproofstyle}

\declaretheorem[numberlike=equation]{observation}

\declaretheorem[name=Observation,numbered=no]{observation*}

\declaretheorem[numberlike=equation]{theorem}

\declaretheorem[name=Theorem,numbered=no]{theorem*}

\declaretheorem[numberlike=equation]{lemma}
\declaretheorem[name=Lemma,numbered=no]{lemma*}

\declaretheorem[name=Corollary,numbered=no]{corollary*}

\declaretheorem[name=Proposition,numbered=no]{proposition*}

\declaretheorem[numberlike=equation]{claim}
\declaretheorem[name=Claim,numbered=no]{claim*}

\declaretheorem[numberlike=equation]{conjecture}
\declaretheorem[name=Conjecture,numbered=no]{conjecture*}

\declaretheorem[name=Question,numbered=no]{question*}

\declaretheoremstyle[bodyfont=\it,qed=$\lozenge$]{defstyle} 

\declaretheorem[numberlike=equation,style=defstyle]{definition}
\declaretheorem[unnumbered,name=Definition,style=defstyle]{definition*}

\declaretheorem[unnumbered,name=Example,style=defstyle]{example*}

\declaretheorem[unnumbered,name=Notation=defstyle]{notation*}

\declaretheorem[unnumbered,name=Construction,style=defstyle]{construction*}

\declaretheorem[numberlike=equation,style=defstyle]{remark}
\declaretheorem[unnumbered,name=Remark,style=defstyle]{remark*}

\newif\ifdraft
\drafttrue 

\ifdraft

\newcommand{\PCnote}[1]{\textcolor{BrickRed}{\guillemotleft PC: #1 \guillemotright}}
\newcommand{\ATnote}[1]{\textcolor{OliveGreen}{\guillemotleft AT: #1 \guillemotright}}
\newcommand{\KGnote}[1]{\textcolor{MidnightBlue}{\guillemotleft KG: #1 \guillemotright}}
\newcommand{\pending}[1]{#1 \vspace{1em}}
\else

\newcommand{\PCnote}[1]{}
\newcommand{\ATnote}[1]{}
\newcommand{\KGnote}[1]{}
\newcommand{\pending}[1]{}
\fi

\newcommand{\ignore}[1]{}

%% file: notationMacros.tex
\newcommand{\inparen}[1]{\left( #1 \right)}

\newcommand{\inbrace}[1]{\left\{ #1 \right\}}

\newcommand{\inangle}[1]{\left< #1 \right>}

\newcommand{\set}[1]{\inbrace{#1}}

\newcommand{\abs}[1]{\left| #1 \right|}

\newcommand{\coeff}{\operatorname{coeff}}
\newcommand{\size}{\operatorname{size}}
\newcommand{\supp}{\operatorname{supp}}

\newcommand{\VP}{\mathsf{VP}}
\newcommand{\VNP}{\mathsf{VNP}}
\newcommand{\VPSPACE}{\mathsf{VPSPACE}}
\def\P{\mathsf{P}}
\newcommand{\NP}{\mathsf{NP}}
\newcommand{\PSPACE}{\mathsf{PSPACE}}

\newcommand{\Det}{\operatorname{\sf Det}}
\newcommand{\Perm}{\operatorname{\sf Perm}}

\def\phi{\varphi}   
\def\epsilon{\varepsilon}   

%% file: localMacros.tex
\newcommand{\mVPSPACE}{\mathsf{mVPSPACE}}
\newcommand{\mVNP}{\mathsf{mVNP}}
\newcommand{\mVP}{\mathsf{mVP}}

\newcommand{\polytope}{\mathsf{Newt}}
\newcommand{\suppy}{\operatorname{supp}_{\vecy}}

\newcommand{\project}[2]{\mathtt{fix}_{(#1 = #2)}}

\newcommand{\summation}{\mathtt{sum}}
\newcommand{\production}{\mathtt{prod}}
\newcommand{\quant}{\mathtt{Q}}
\newcommand{\calA}{\mathcal{A}}
\newcommand{\calB}{\mathcal{B}}
\newcommand{\calP}{\mathcal{P}}
\newcommand{\ckt}{\mathcal{C}}

\newcommand{\Thomasse}{Thomass\'{e}}

%% file: customurlbst/bibmacros.tex
\usepackage{nth}
\usepackage{intcalc}
\usepackage{etoolbox}
\usepackage{xstring}

\usepackage{ifpdf}
\ifpdf
\else
\usepackage[quadpoints=false]{hypdvips}
\fi

\newcommand{\ECCCurl}[2]{http://eccc.hpi-web.de/report/\ifnumcomp{#1}{>}{93}{19}{20}#1/#2/}

\newcommand{\shortECCC}[2]{\texttt{\href{http://eccc.hpi-web.de/report/\ifnumcomp{#1}{>}{93}{19}{20}#1/#2/}{eccc:TR#1-#2}}}

\newcommand{\parseECCC}[1]{
\StrSubstitute{#1}{TR}{}[\tmpstring]%
\IfSubStr{\tmpstring}{/}{ 
\StrBefore{\tmpstring}{/}[\ecccyear]%
\StrBehind{\tmpstring}{/}[\ecccreport]%
}{
\StrBefore{\tmpstring}{-}[\ecccyear]%
\StrBehind{\tmpstring}{-}[\ecccreport]%
}%
\shortECCC{\ecccyear}{\ecccreport}}

%% file: main.bbl
\begin{thebibliography}{CDGM22}

\bibitem[CDGM22]{CDGM22}
Arkadev Chattopadhyay, Rajit Datta, Utsab Ghosal, and Partha Mukhopadhyay.
\newblock \href {http://dx.doi.org/10.4230/LIPIcs.ITCS.2022.39} {Monotone
  Complexity of Spanning Tree Polynomial Re-Visited}.
\newblock In {\em 13th Innovations in Theoretical Computer Science Conference,
  {ITCS} 2022, January 31 - February 3, 2022, Berkeley, CA, {USA}}, volume 215
  of {\em LIPIcs}, pages 39:1--39:21. Schloss Dagstuhl - Leibniz-Zentrum
  f{\"{u}}r Informatik, 2022.

\bibitem[CDM21]{CDM21}
Arkadev Chattopadhyay, Rajit Datta, and Partha Mukhopadhyay.
\newblock \href {http://dx.doi.org/10.1145/3406325.3451069} {Lower bounds for
  monotone arithmetic circuits via communication complexity}.
\newblock In {\em {STOC} '21: 53rd Annual {ACM} {SIGACT} Symposium on Theory of
  Computing, Virtual Event, Italy, June 21-25, 2021}, pages 786--799. {ACM},
  2021.

\bibitem[CGM22]{CGM22}
Arkadev Chattopadhyay, Utsab Ghosal, and Partha Mukhopadhyay.
\newblock \href {http://dx.doi.org/10.4230/LIPIcs.FSTTCS.2022.12} {Robustly
  Separating the Arithmetic Monotone Hierarchy via Graph Inner-Product}.
\newblock In {\em 42nd {IARCS} Annual Conference on Foundations of Software
  Technology and Theoretical Computer Science, {FSTTCS} 2022, December 18-20,
  2022, {IIT} Madras, Chennai, India}, volume 250 of {\em LIPIcs}, pages
  12:1--12:20. Schloss Dagstuhl - Leibniz-Zentrum f{\"{u}}r Informatik, 2022.

\bibitem[CKR20]{CKR20}
Bruno~Pasqualotto Cavalar, Mrinal Kumar, and Benjamin Rossman.
\newblock \href {http://dx.doi.org/10.1007/978-3-030-61792-9\_25} {Monotone
  Circuit Lower Bounds from Robust Sunflowers}.
\newblock In {\em {LATIN} 2020: Theoretical Informatics - 14th Latin American
  Symposium, S{\~{a}}o Paulo, Brazil, January 5-8, 2021, Proceedings}, volume
  12118 of {\em Lecture Notes in Computer Science}, pages 311--322. Springer,
  2020.

\bibitem[Gas87]{Gas87}
S.B. Gashkov.
\newblock The complexity of monotone computations of polynomials.
\newblock {\em Moscow University Math Bulletin}, (5):1–8, 1987.

\bibitem[Gro17]{G17}
Joshua~A. Grochow.
\newblock \href {http://dx.doi.org/10.4086/toc.2017.v013a018} {Monotone
  Projection Lower Bounds from Extended Formulation Lower Bounds}.
\newblock {\em Theory of Computing}, 13(18):1--15, 2017.

\bibitem[GS12]{GS12}
S.~B. Gashkov and I.~S. Sergeev.
\newblock A method for deriving lower bounds for the complexity of monotone
  arithmetic circuits computing real polynomials.
\newblock {\em Sbornik. Mathematics}, 203(10), 2012.

\bibitem[HY13]{HY13}
Pavel Hrube\v{s} and Amir Yehudayoff.
\newblock \href {http://dx.doi.org/10.1109/CCC.2013.11} {Formulas are
  Exponentially Stronger than Monotone Circuits in Non-commutative Setting}.
\newblock In {\em Proceedings of the 28th Conference on Computational
  Complexity, {CCC} 2013, K.lo Alto, California, USA, 5-7 June, 2013}, pages
  10--14. {IEEE} Computer Society, 2013.

\bibitem[HY16]{HY16}
Pavel Hrubes and Amir Yehudayoff.
\newblock \href {http://dx.doi.org/10.4230/LIPIcs.ICALP.2016.89} {On
  Isoperimetric Profiles and Computational Complexity}.
\newblock In {\em 43rd International Colloquium on Automata, Languages, and
  Programming, {ICALP} 2016, July 11-15, 2016, Rome, Italy}, volume~55 of {\em
  LIPIcs}, pages 89:1--89:12. Schloss Dagstuhl - Leibniz-Zentrum f{\"{u}}r
  Informatik, 2016.

\bibitem[HY21]{HY21}
\mfbiberr{toupdate(HY21): Journal}Pavel Hrube\v{s} and Amir Yehudayoff.
\newblock \href {http://dx.doi.org/10.4230/LIPIcs.CCC.2021.9} {Shadows of
  Newton Polytopes}.
\newblock In {\em 36th Computational Complexity Conference, {CCC} 2021, July
  20-23, 2021, Toronto, Ontario, Canada (Virtual Conference)}, volume 200 of
  {\em LIPIcs}, pages 9:1--9:23. Schloss Dagstuhl - Leibniz-Zentrum f{\"{u}}r
  Informatik, 2021.

\bibitem[JS82]{JS82}
Mark Jerrum and Marc Snir.
\newblock \href {http://dx.doi.org/10.1145/322326.322341} {Some Exact
  Complexity Results for Straight-Line Computations over Semirings}.
\newblock {\em Journal of the ACM}, 29(3):874--897, 1982.

\bibitem[KP09a]{KP09b}
Pascal Koiran and Sylvain Perifel.
\newblock \href {http://dx.doi.org/10.1016/j.tcs.2009.08.026} {{VPSPACE} and a
  transfer theorem over the complex field}.
\newblock {\em Theor. Comput. Sci.}, 410(50):5244--5251, 2009.

\bibitem[KP09b]{KP09}
Pascal Koiran and Sylvain Perifel.
\newblock \href {http://dx.doi.org/10.1007/s00037-009-0269-1} {{VPSPACE} and a
  Transfer Theorem over the Reals}.
\newblock {\em Computational Complexity}, 18(4):551--575, 2009.

\bibitem[KPTT15]{KPTT15}
Pascal Koiran, Natacha Portier, S\'{e}bastien Tavenas, and St\'{e}phan
  Thomass\'{e}.
\newblock \href {http://dx.doi.org/10.1007/s10208-014-9216-x} {A
  $\tau$-Conjecture for Newton Polygons}.
\newblock {\em Foundations of Computational Mathematics}, 15:185--197, 2015.

\bibitem[KZ86]{KZ86}
O.~M. Kasim-Zade.
\newblock Arithmetic complexity of monotone polynomials.
\newblock {\em Theoretical Problems in Cybernetics. Abstracts of lectures},
  page 68–69, 1986.

\bibitem[Mal11]{M11}
Guillaume Malod.
\newblock \href {http://dx.doi.org/10.1007/978-3-642-22953-4\_18} {Succinct
  Algebraic Branching Programs Characterizing Non-uniform Complexity Classes}.
\newblock In {\em Fundamentals of Computation Theory - 18th International
  Symposium, {FCT} 2011, Oslo, Norway, August 22-25, 2011. Proceedings}, pages
  205--216, 2011.

\bibitem[MR13]{MR13}
Meena Mahajan and B.~V.~Raghavendra Rao.
\newblock \href {http://dx.doi.org/10.1007/s00037-011-0024-2} {Small Space
  Analogues of Valiant's Classes and the Limitations of Skew Formulas}.
\newblock {\em Computational Complexity}, 22(1):1--38, 2013.

\bibitem[Poi08]{P08}
Bruno Poizat.
\newblock \href {https://doi.org/10.2178/jsl/1230396913} {A la recherche de la
  definition de la complexite d'espace pour le calcul des polynomes a la
  maniere de Valiant}.
\newblock {\em J. Symb. Log.}, 73(4):1179--1201, 2008.

\bibitem[RY11]{RY11}
Ran Raz and Amir Yehudayoff.
\newblock \href {http://dx.doi.org/10.1016/j.jcss.2010.06.013} {Multilinear
  formulas, maximal-partition discrepancy and mixed-sources extractors}.
\newblock {\em Journal of Computer and System Sciences}, 77(1):167--190, 2011.

\bibitem[Sch76]{Sch76}
Claus{-}Peter Schnorr.
\newblock \href {http://dx.doi.org/10.1016/0304-3975(76)90083-9} {A Lower Bound
  on the Number of Additions in Monotone Computations}.
\newblock {\em Theor. Comput. Sci.}, 2(3):305--315, 1976.

\bibitem[Sri20]{S20}
Srikanth Srinivasan.
\newblock \href {http://dx.doi.org/10.1145/3417758} {Strongly Exponential
  Separation between Monotone {VP} and Monotone {VNP}}.
\newblock {\em {ACM} Trans. Comput. Theory}, 12(4):23:1--23:12, 2020.

\bibitem[SS77]{SS77}
Eli Shamir and Marc Snir.
\newblock {\em Lower bounds on the number of multiplications and the number of
  additions in monotone computations}.
\newblock IBM Thomas J. Watson Research Division, 1977.

\bibitem[SS80]{SS80}
Eli Shamir and Marc Snir.
\newblock \href {http://dx.doi.org/10.1007/BF01744302} {On the Depth Complexity
  of Formulas}.
\newblock {\em Math. Syst. Theory}, 13:301--322, 1980.

\bibitem[Val80]{V80}
Leslie~G. Valiant.
\newblock \href {http://dx.doi.org/10.1016/0304-3975(80)90060-2} {Negation can
  be Exponentially Powerful}.
\newblock {\em Theor. Comput. Sci.}, 12:303--314, 1980.

\bibitem[Yeh19]{Y19}
Amir Yehudayoff.
\newblock \href {http://dx.doi.org/10.1145/3313276.3316311} {Separating
  monotone {VP} and {VNP}}.
\newblock In {\em Proceedings of the 51st Annual {ACM} {SIGACT} Symposium on
  Theory of Computing, {STOC} 2019, Phoenix, AZ, USA, June 23-26, 2019}, pages
  425--429. {ACM}, 2019.

\end{thebibliography}
